# Challenges and Research Directions for Large Language Model Inference Hardware

Xiaoyu Ma and David Patterson, Google

We highlight four promising research opportunities to improve *Large Language Model* inference for datacenter AI: *High Bandwidth Flash* for 10X memory capacity with HBM-like bandwidth; Processing-Near-Memory and 3D memory-logic stacking for high memory bandwidth; and low-latency interconnect to speedup communication. We also review their applicability for mobile devices.

## INTRODUCTION

When one author started his career in 1976, ~40% of the papers at computer architecture conferences were from industry. Their share fell below 4% at the 2025 International Symposium on Compute Architecture, suggesting a near disconnect between research and practice. To help restore their historic bond, we propose research directions that, if pursued, address some of the biggest hardware challenges that the AI industry faces.

**Large language model (LLM) inference is a crisis.** Rapidly improving hardware enables AI advances. Projections of inference chip annual sales are 4X-6X over the next 5-8 years.[1] While training demonstrates remarkable AI breakthroughs, the cost of inference determines economic viability. Companies find it costly to serve state-of-the-art models as usage of these models dramatically increases.[2,3]

**New trends make inference harder.** Recent advances in LLMs require more resources for inference:

- **Mixture of Experts (MoE)**. Rather than a single dense feedforward block, MoE uses tens to hundreds of experts—256 for DeepSeekv3— invoked selectively. This sparsity allows model size to grow significantly for higher quality, despite a modest increase in training cost. While helping training, MoE exacerbates inference by expanding memory and communication.
- **Reasoning models**. Reasoning is a think-before-act technique to improve quality. An extra "thinking" step generates a long sequence of "thoughts" before the final answer, similar to people solving a problem step-by-step. Thinking greatly increases generation latency, and the long sequence of thought tokens strains memory.
- **Multimodal**. LLMs have evolved from text to image, audio, and video generation. Larger data types demand more than text generation.
- **Long context.** A context window refers to the amount of information the LLM model can look at when generating an answer. Longer context helps quality, but increases compute and memory demands.
- **Retrieval-Augmented Generation (RAG**). RAG accesses a user-specific knowledge database to obtain relevant information as extra context to improve LLM results, increasing resource demands.
- **Diffusion.** In contrast to the autoregressive method that generates tokens sequentially, the novel diffusion method generates all tokens (e.g., an entire image) in one step and then iteratively denoises the image to reach desired quality. Unlike above, diffusion only increases compute demands.

The growing market and challenges of LLM inference suggest a great opportunity and need for innovation!

# CURRENT LLM INFERENCE HARDWARE AND ITS INEFFICIENCIES

We first review LLM inference basics and its primary bottlenecks on mainstream AI architectures, focusing on LLMs in the datacenter. LLMs on mobile devices have different restrictions and thus different options (e.g., HBM is infeasible).

LLMs, whose heart is Transformer, have two inference phases with very different characteristics: *Prefill* and *Decode* (Figure 1). Prefill is similar to training by processing all tokens of the input sequence simultaneously, so it is inherently parallel and often compute bound. In contrast, Decode is inherently sequential, as each step generates one output token ("autoregressive"), making it memory bound. The *Key Value* (*KV) Cache* connects the two phases, with its size proportional to the input+output sequence length. Although together in Figure 1, Prefill and Decode are not tightly coupled, and often run on different servers. Disaggregated inference allows software optimizations like batching to make Decode be less memory bound. A survey for efficient LLM inference reviews many software optimizations.[4]

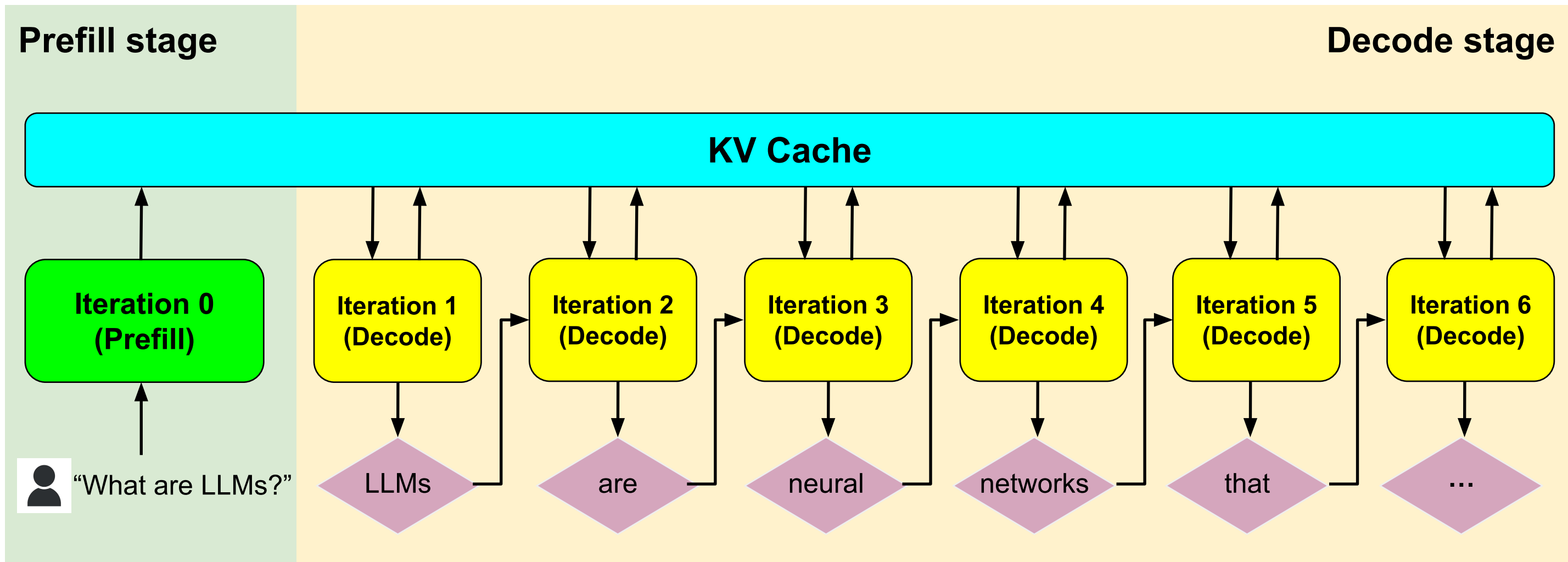

**Figure 1. The key processes of inference for the Transformer model that is the foundation of LLMs.**

GPUs and Google's Tensor Processing Units (TPUs) are popular datacenter accelerators for both training and inference. Historically, inference versions were scaled-down from training systems, such as with fewer chips or a smaller chip with less memory or performance. Thus far, no GPU/TPU was designed solely for LLM inference. Because Prefill is similar to training whereas Decode differs significantly, two challenges make GPUs/TPUs inefficient for Decode.

## Decode Challenge #1: Memory

The autoregressive Decode makes inference inherently memory bound, with new software trends heightening this challenge. In contrast, the hardware trends go in a completely different direction.

**AI processors face a Memory Wall.** Current datacenter GPUs/TPUs rely on *High Bandwidth Memory* (*HBM*), and connect several HBM stacks to a single monolithic accelerator application specific integrated circuit (ASIC) (see Figure 2 and Table 1). Nevertheless, memory bandwidth improves more slowly than compute floating-point operations per second (FLOPS). For example, NVIDIA GPU 64-bit FLOPS rose by 80X from 2012 to 2022, but bandwidth grew only 17X. This gap will continue expanding.

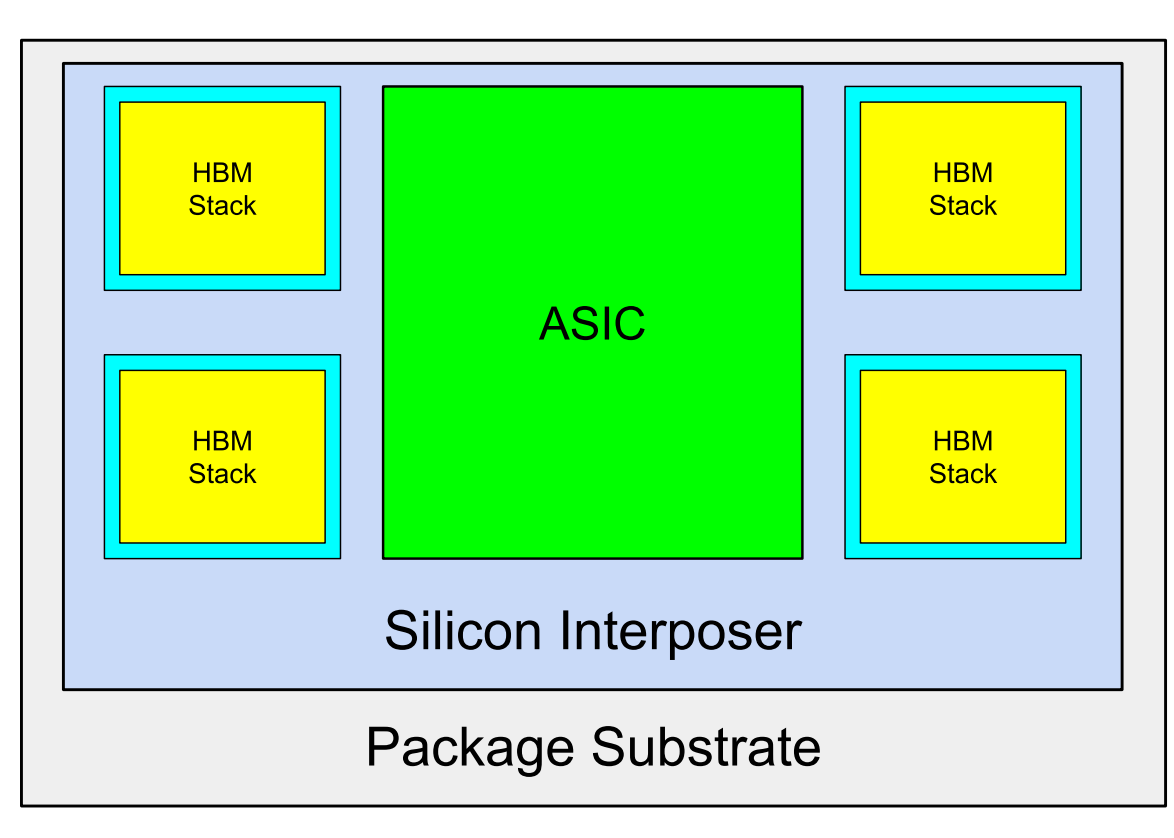

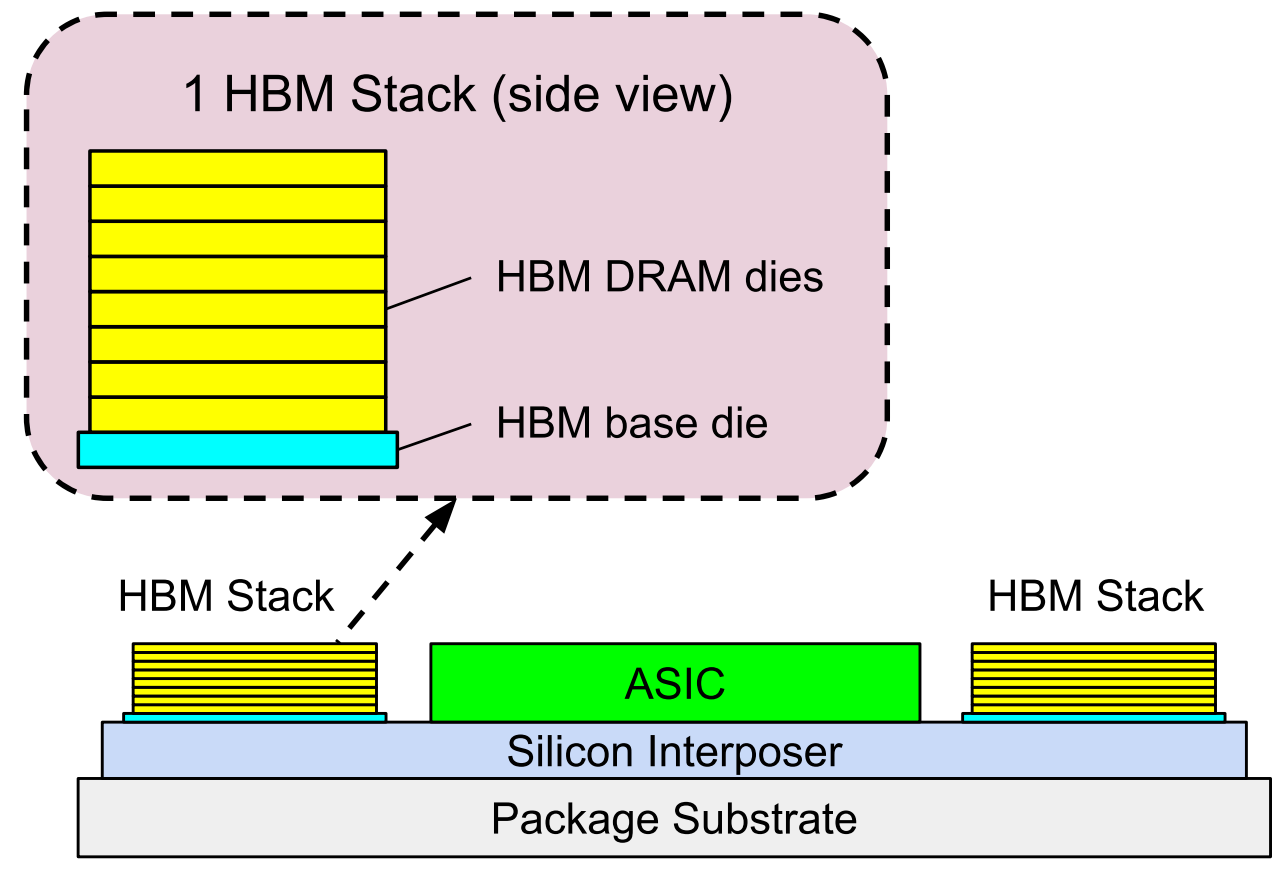

**(a) HBM (Top View)** **(b) HBM (Side View)**

**Figure 2. (a) High Bandwidth Memory (HBM) package top view, (b) HBM side view.**

| | HBM | HBM2 | HBM2E | HBM3 | HBM3E | HBM4 |
|---|---|---|---|---|---|---|
| Year Introduced | 2013 | 2016 | 2019 | 2022 | 2023 | 2026 |
| Max pin bandwidth (gigabit/sec) | 1.0 | 2.4 | 3.6 | 6.4 | 9.8 | 8 |
| Number of pins | 1024 | 1024 | 1024 | 1024 | 1024 | 2048 |
| Stack BW (gigabyte/sec) | 128 | 307 | 461 | 819 | 1254 | 2048 |
| Max Number of dies/Stack | 4 | 8 | 12 | 12 | 16 | 16 |
| Max Capacity per die (gibibyte) | 1 | 1 | 2 | 2 | 3 | 4 |
| Max Stack Capacity (gibibyte) | 4 | 8 | 24 | 24 | 48 | 64 |
| NVIDIA GPU Generation | | Volta V100 | Ampere A100 | Hopper H100 | Blackwell B100 | Rubin R100 |
| HBM stacks/GPU | | 4 | 5 | 5 | 8 | 8 |

**Table 1. Key features of six generations of HBM.**

**HBM is increasingly expensive.** Looking at one HBM stack, the normalized price of capacity ($/GB) and bandwidth ($/GBps) *increases* over time. Figure 3(a) shows both grew 1.35x higher from 2023-2025.[5] This rise is because manufacturing and packaging difficulties increase with dies per HBM stack and Dynamic RAM (DRAM) density growth. In contrast, Figure 3(b) shows the equivalent costs for standard Double Data Rate (DDR4) DRAM *decrease* over time. From 2022-2025, capacity cost shrank to 0.54x and bandwidth cost to 0.45x. While prices of all memory and storage devices surged in 2026 due to unexpected demand, we believe long term that the diverging pricing trends of HBM and DRAM will hold.

**DRAM density growth is decelerating.** For an individual DRAM die, scaling is also worrisome. Fourfold growth from 8-gigabit DRAM dies that debuted in 2014 will take over 10 years. Fourfold gains occurred every 3-6 years previously.

**SRAM-only solutions are insufficient**. Cerebras and Groq tried using full reticle chips filled with SRAM to avoid DRAM and HBM challenges. (Cerebras even used wafer scale integration.) While plausible when the companies were founded a decade ago, LLMs soon overwhelmed on-chip SRAM capacity. Both had to later retrofit external DRAM.

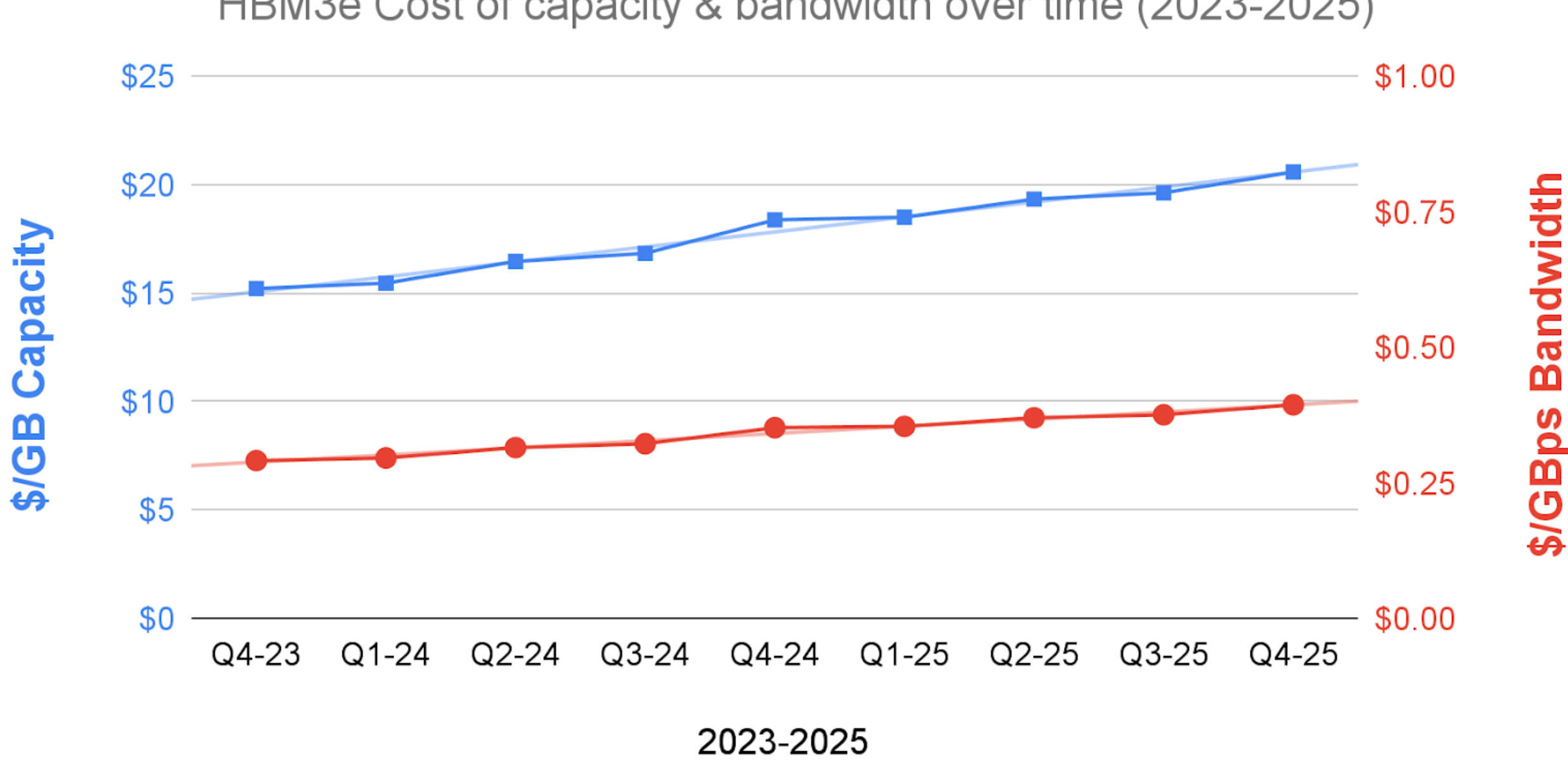

(a) HBM increasing $/GB capacity and $/GBps bandwidth.

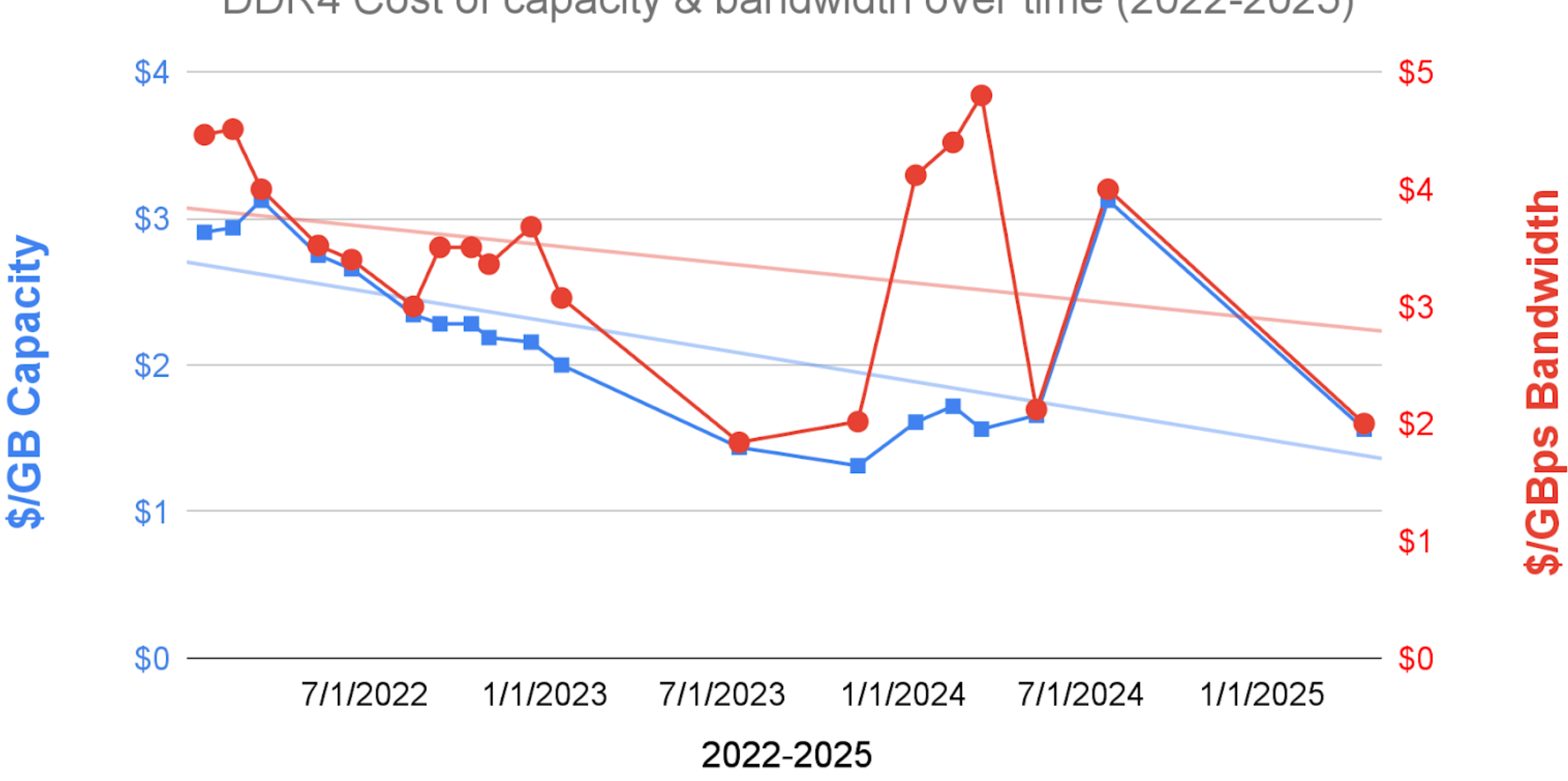

(b) DDR decreasing $/GB capacity and $/GBps bandwidth (Source: see Appendix)

**Figure 3. Cost per capacity and Bandwidth over time with trendlines for HBM (a) vs DDR (b).**

## Decode Challenge #2: End-to-End Latency

**User-facing implies low latency.** Unlike training that takes weeks, inference is tied to real-time requests, needing a response in seconds or less. Low latency is critical for user-facing inference. (Batch or offline inference does not have a low latency requirement.) Depending on the application, latency is measured as *time-to-completion* of all output tokens or *time-to-first-token*. Both have challenges:

- **Time-to-completion challenge**. Decode produces one token at a time, so the longer the output, the longer the latency. Long output sequences stretch latency, but long input sequences are also slower because accessing the KV Cache during Decode and Prefill takes more time. Each Decode iteration has high memory access latency because it is memory bound.
- **Time-to-first-token challenge.** Long input sequences and RAG increase the amount of work before generation and hence the time-to-first-token. Reasoning models also increase this latency as they generate many “thought” tokens before the first user-visible token.

**Interconnect latency outweighs bandwidth.** Before LLMs, in datacenter inference usually ran on one chip, while training needed a supercomputer. The supercomputer interconnect understandably aimed much more at bandwidth than latency. LLM inference changes the game:
- Because of big weights, LLM inference now needs a multi-chip system, with software sharding that implies frequent communication. MoE and long sequence models further increase the system size to accommodate larger memory capacity requirements.
- Unlike training, the size of network messages is often small, given the small batch size of Decode. Latency trumps bandwidth for frequent, small messages in a big network.

Table 2 summarizes the main challenges of Decode inference. Only Diffusion needs increased compute—relatively easy to deliver—as it is fundamentally unlike Transformer Decode. Thus, we focus on promising directions for improving memory and interconnect latency but not compute. The last four rows are research opportunities to fulfill these needs, covered next.

| | **LLM Decode features & trends + Promising research opportunities** | **Memory capacity** | **Memory bandwidth** | **Interconnect latency** | **Compute** |
|---|---|---|---|---|---|
| Drivers of hardware improvements | Conventional Transformer Decode | | ✔ | ✔ | |
| | MoE | ✔ | ✔ | ✔ | |
| | Reasoning models | ✔ | ✔ | ? | |
| | Multimodal | ✔ | ✔ | ? | |
| | Long-context | ✔ | ✔ | ? | |
| | RAG | ✔ | ✔ | ? | |
| | Diffusion | | | | ✔ |
| Where promising directions can help | ① High Bandwidth Flash | ✔ | | ⇧ | |
| | ② Near Memory Compute | | ✔ | ⇧ | |
| | ③ 3D Compute-Logic Stacking | | ✔ | ⇧ | |
| | ④ Low-Latency Interconnect | | | ✔ | |

**Table 2. Summary of primary hardware bottlenecks for LLM inference and research directions to address them.** “✔” means primary bottlenecks. In the top section of the table (hardware improvement drivers), “?” means a derived interconnect bottleneck for the drivers. For example, if reasoning models require a larger system to fulfill the memory requirements, it puts pressure on interconnect latency by increasing the hop count. Likewise, if an improvement on memory capacity or bandwidth allows inference of the same model with fewer accelerator chips, it would help lower interconnect latency. For the bottom section (promising directions), “⇧” means a promising direction that helps with interconnect latency by shrinking the size of the overall system, thereby reducing the hop count.

## FOUR RESEARCH OPPORTUNITIES TO RE-THINK LLM INFERENCE HARDWARE

Performance/cost metrics measure the efficiency of AI systems. Modern metrics—which emphasize realistic performance normalized, *Total Cost of Ownership (TCO)*, average power consumption, and *carbon dioxide equivalent emissions* ($CO_2e$)—offer new targets for designing systems:[6]

- **Performance must be meaningful**. For LLM Decode inference, high FLOPS on a giant die does not necessarily mean high performance. Instead, we need to scale memory bandwidth and capacity efficiently, and optimize interconnect speed.
- **Performance must be delivered within datacenter capacity**, often constrained by power, space, and $CO_2e$ budgets.
- **Power and $CO_2e$ are first-order optimization targets**. Power affects TCO and datacenter capacity. Power and energy cleanliness determine operational $CO_2e$. Manufacturing yield and life cycle set embodied $CO_2e$.

Next, we describe four promising research directions to address the Decode challenges (bottom of Table 2). Although described independently, they are synergistic; an architecture can usefully combine many of them. All improve performance/TCO, performance/$CO_2e$, and performance/power.

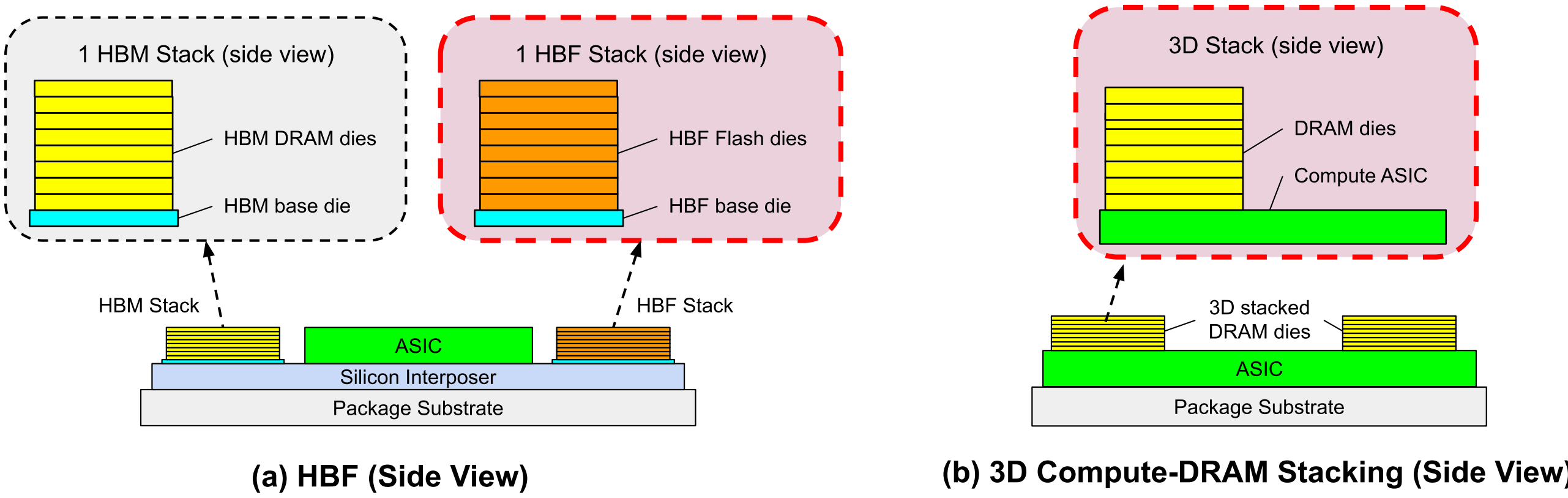

**Figure 4. (a) High Bandwidth Flash (HBF) and (b) 3D Compute-logic Stacking**

### ① High Bandwidth Flash for 10X capacity

*High Bandwidth Flash* (*HBF*) combines HBM bandwidth with flash capacity by stacking flash dies like HBM (Figure 4 (a)).[7] HBF delivers 10X memory capacity per node to reduce system size to save power, TCO, CO2e, and network overhead. Table 3 compares HBF to HBM and DDR and low power DDR (LPDDR) DRAM. The weaknesses of alternatives are bandwidth for DDR5, capacity for HBM, and write limits and high read latency for HBF. Another HBF benefit is sustainable capacity scaling; flash capacity continues to double every three years while, as mentioned above, DRAM growth decelerates.

Two well-known flash memory restrictions must be addressed:

- **Limited write endurance**. Write/erase cycles can wear-out flash. Therefore, HBF must hold infrequently-updated data, such as weights at inference time or slow-changing context.
- **Page-based reads with high latency**. Flash reads are at page granularity (10s KBs) with a latency substantially worse than DRAM (microseconds). Small reads reduce effective bandwidth.

These issues mean HBF cannot replace all HBM; a system still needs normal DRAM for data unsuitable for HBF.

| | Capacity (GB) | Bandwidth (GB/s) | Power (Watt) | GBps per Watt | GB per Watt | Read latency (ns) | Bytes per read | Write endurance |
|---|---|---|---|---|---|---|---|---|
| 1 HBF stack | 512 | 1638 (read) | <80 | >20.5 | >6.4 | 1,000s | 4096 | low |
| 1 HBM4-6400 stack | 48 | 1638 | 40 | 41 | 1 | 10-100 | 32 | high |
| 1 DDR5-6400-64GB module | 64 | 51 | 12 | 4 | 5 | 10-100 | 64 | high |
| 1 LPDDR5-6400-16GB module | 16 | 51 | 3 | 17 | 5 | 10-100 | 64 | high |
| 1 Flash card | 4096 | 4 (read) | 50 | 0.1 | 82 | 10,000s | 4096 | low |

**Table 3. A ballpark comparison of HBF, HBM, DDR, LPDDR, and flash.**

The addition of HBF enables exciting capabilities for LLM inference:

- **10X weight memory**. Weights are frozen during inference, so HBF 's 10X capacity could host many more weights—such as giant MoEs—to enable much bigger models than affordable today.
- **10X context memory.** Limited write endurance makes HBF infeasible for KV Cache data updated for every query or generation token. However, it works for a slow-changing context. For example:
  - A web corpus, used by LLM search, that stores billions of Internet documents.
  - A code database, used by AI coding, that stores billions of lines of code.
  - A paper corpus, used by AI tutoring, that tracks millions of research papers.
- **Smaller inference system**. Memory capacity determines the minimum hardware to hold a model. HBF downsizes the system, helping communication, reliability, and resource allocation.
- **Greater resource capacity.** HBF would reduce dependency on HBM-only architectures and alleviate the global shortage of mainstream memory devices.

HBF opens new research questions:

- How can software deal with limited write endurance and page-based, high-latency reads?
- What should be the ratio of traditional memory to HBF in a system?
- Can we reduce the constraints of HBF technology itself?
- How should HBF be configured for mobile devices versus for datacenters?

## ② Processing-Near-Memory for high bandwidth

*Processing-in-Memory (PIM),* conceived several decades ago[8], obtains high bandwidth by augmenting memory dies with small, low-power processors attached to memory banks. While PIM offers extraordinary bandwidth, key challenges are software sharding and memory-logic coupling. The former limits the number of software kernels that can run well on PIM. The latter hurts power and area efficiency of compute logic. In contrast, *Processing-Near-Memory (PNM) is* a technique that places memory and logic nearby but still uses separate dies. One version of PNM is 3D compute-logic stacking (see ③).

Unfortunately, some recent papers blur the distinction between PIM and PNM. They use PIM as a general term whether or not the compute logic is placed directly into the memory die. We go here with a simple but sharp distinction: PIM refers to designs where the processor and memory are in the same die and PNM means they are on nearby but separate dies. This distinction makes PIM and PNM unambiguous.

Hardware advantages are irrelevant if it's too hard for software to use, which is our experience for PIM and datacenter LLMs. Table 4 lists why PNM is better than PIM for LLM inference, despite weaknesses in bandwidth and power. Specifically, PIM requires software to shard memory structures of LLMs into many small pieces that rarely interact to fit into 32-64MB memory banks; shards in PNM can be 1000x larger, making it much easier to partition LLMs with a low communication overhead. It is also unclear if the compute can be sufficient in PIM given the very limited budget for power and thermal of a DRAM technology process node.

| | Processing-in-Memory (PIM) | Processing-Near-Memory (PNM) | Winner |
|---|---|---|---|
| Examples | Samsung HBM-PIM[9]<br>SK Hynix GDDR-PIM[10]<br>UPMEM logic die on DIMM[11] | Compute on HBM base die[12,13]<br>AMD DRAM-logic 3D stacking[14]<br>Marvell Structera CXL-PNM[15] | — |
| Data movement power | Very low (on-chip) | Low (off-chip but nearby) | PIM |
| Bandwidth (per Watt) | Very high (5X-10X of standard) | High (2X-5X of standard) | PIM |
| Memory-logic coupling | Memory and logic on one die | Memory and logic on separate dies | PNM |
| Logic PPA (performance, power, area) | Slower and higher-power logic if in a DRAM process | Logic in a logic process helps performance, power, and area. | PNM |
| Memory density | Worse since shared with logic | Not affected | PNM |
| Commodity memory pricing (per GB) | No. Lower volume, fewer suppliers, lower density vs memory w/o logic | Yes. Not affected | PNM |
| Power/Thermal budget | Logic has tight power and thermal budget on a memory die | Logic is less constrained by power and thermal limits | PNM |
| Software sharding | Bank parallelism needs sharding workloads to banks (e.g., 32–64 MB) | Less restrictive on sharding (e.g., 16–32 GB). No need to shard to memory banks | PNM |

**Table 4. PIM versus PNM for datacenter LLM inference.**

While PNM is better than PIM for datacenter LLMs, the comparison is not as clear for mobile devices. Mobile devices are more energy-constrained and run LLMs with many fewer weights, shorter context, smaller data types, and smaller batch sizes due to a single user. These differences simplify sharding, reduce the compute and thermal need, making PIM weaknesses less problematic and so plausibly viable for mobile devices.

### ③ 3D memory-logic stacking for high bandwidth

Unlike 2D hardware, where memory IOs reside on the shoreline, 3D stacking (see Figure 4(b)) instead uses vertical *through silicon vias* (*TSVs*) to get a wide-and-dense memory interface for high bandwidth at low power.

3D memory-logic stacking has two versions:

1. **Compute-on-HBM-base-die** reuses HBM designs by inserting the compute logic into the HBM base die.[12,13] Because the memory interface is unchanged, bandwidth is the same as HBM, while power is 2–3X lower because of the shortened data path.
2. **Custom 3D** solutions enable bandwidth and bandwidth-per-watt higher than reusing HBM through using a wider-and-denser memory interface and more advanced packaging technologies.

Despite better bandwidth and power, 3D stacking faces challenges:

1. **Thermal.** Cooling a 3D design is harder than 2D as there is less surface area. One solution is to limit FLOPS of the compute logic by running at a low clock speed and voltage, as LLM Decode inference already has low arithmetic intensity.

2. **Memory-logic coupling.** An industry standard may be required for the memory interface of 3D compute-logic stacking.

3D stacking opens new research questions:
- The ratio of memory bandwidth to capacity or compute FLOPS is significantly different from existing systems. How can software adapt to it?
- Imagine a system with many memory types. How do we map LLMs efficiently?
- How to communicate with other memory-logic stacks and the main AI processor (if necessary)?
- What are tradeoffs in bandwidth, power, thermal, and reliability for various design choices, e.g., compute die placed on top vs bottom, the memory die count per stack, …?
- How do these opportunities change for mobile devices versus datacenter LLM accelerators?

### ④ Low-latency interconnect

Techniques ①–③ help latency as well as throughput: higher memory bandwidth reduces latency of every Decode iteration and higher memory capacity per accelerator chip reduces system size, saving communication overhead. Another promising latency direction for datacenters is to rethink the network latency-bandwidth tradeoff since inference is more sensitive to interconnect latency. For example:
- **High-connectivity topology.** Topologies with high connectivity—such as tree, dragonfly, and high-dimensional tori—require fewer hops, reducing latency. These topologies may diminish bandwidth but improve latency.
- **Processing-in-network.** Communication collectives used by LLMs—broadcast, all-reduce, MoE dispatch and collect—are well suited for in-network acceleration to improve both bandwidth and latency, e.g., a tree topology with in-network aggregation enables both low-latency and high-throughput all-reduce.
- **AI chip optimization.** The latency focus influences chip design with several possible optimizations:
  - Storing small arriving packets directly into on-chip SRAM instead of off-chip DRAM;
  - Placing the compute engine close to the network interface to reduce transportation time.
- **Reliability.** Codesigning reliability and interconnect can help both:
  - A local standby spare reduces system failures and the latency and throughput consequences of migrating the failed job to a new healthy node somewhere when there are no standby spares.
  - If perfect communication is unnecessary for LLM inference, one can reduce latency yet deliver satisfactory quality results by using fake data or a prior result when message timeout expires, rather than waiting for straggler messages to arrive.

## RELATED WORK

**High Bandwidth Flash.** SanDisk first proposed HBF, an HBM-like architecture for flash to overcome its bandwidth limit.[7] (SK Hynix later joined its development.) Microsoft researchers proposed a new class of memory that focuses on read performance and high density instead of write performance and retention time for AI inference.[16] While not specifically mentioned, HBF is a concrete example of the proposed new AI memory. Another research paper proposed integrating flash into mobile processors for on-device LLM inference enhanced with an LPDDR interface for the low bandwidth need of Prefill and Processing-Near-Flash for the high bandwidth need of Decode.[17]

**Processing-Near-Memory.** 3D compute-logic stacking has gained increasing attention as a technique for bandwidth higher than HBM, such as compute-on-HBM-base-die proposals[12,13] and an AMD concept[14]. In the non-3D space, Samsung AXDIMM[9] and Marvell Structera-A[15] attach processors to commercial DDR DRAM. The former integrated compute logic in the DIMM buffer chip. The latter leveraged the CXL interface for improved programmability and ease of system integration. (A survey paper provides more examples of PNM/PIM.[18]) Many papers discuss using PIM/PNM in mobile devices, not the main focus of this paper.

**Low-latency interconnect.** Numerous papers describe low hop count network topologies including trees, dragonfly, and high-dimensional Tori. (This magazine's 20 reference limit prevents citation.) Examples of commercial processing-in-network are NVIDIA NVLink and Infiniband switches that support in-switch reduction, and multicast acceleration through SHARP (Scalable Hierarchical Aggregation and Reduction Protocol).[19] Similar capabilities for AI workloads appeared recently in Ethernet switches.[20]

**Software Innovations.** Besides this paper's focus on hardware innovations, there is a rich space of software-hardware codesign for algorithmic and software innovations to improve LLM inference. For example, a root cause is the autoregressive nature of Transformer Decode. A new algorithm that avoided autoregressive generation—such as Diffusion for image generation—could dramatically simplify AI inference hardware.

# CONCLUSION

The increasing importance and difficulty of inference for LLMs—which desperately need lower cost and latency—is an attractive research target. Autoregressive Decode is already a major challenge for memory and interconnect latency, which is exacerbated by MoE, reasoning, multimodal data, RAG, and long input/output sequences.

The computer architecture community has made great contributions on challenges when a realistic simulator was available, as it has previously for branch prediction and cache design. Since primary bottlenecks of LLM inference are memory and latency, a roofline-based performance simulator could be useful to provide first-order estimates in many scenarios. Additionally, such a framework should track memory capacity, explore various sharding techniques that are critical to performance, and use modern performance/cost metrics. We hope academic researchers will respond to this opportunity to accelerate AI research.

The current AI hardware philosophy—full-reticle die with high FLOPS, many HBM stacks, and bandwidth-optimized interconnect—is a mismatch to LLM Decode inference. While many researchers explore compute for datacenters, we recommend instead improving memory and network along four directions: HBF, PNM, 3D stacking, and low latency interconnect. Moreover, novel performance/cost metrics that focus on datacenter capacity, system power, and carbon footprint offer new opportunities versus conventional measures. Constrained versions of HBF, PNM, PIM, and 3D stacking also might work well for mobile device LLMs.

Such advances would unlock collaborative work towards important and urgent innovations that the world needs for delivering affordable AI inference.

# ACKNOWLEDGMENTS

We thank Martin Abadi, Jeff Dean, Norm Jouppi, Amin Vahdat, and Cliff Young for their comments that improved the paper.

## AUTHORS

XIAOYU MA is a Senior Staff Engineer at Google DeepMind, Mountain View, California, 94043. His research interests include domain-specific computer architectures. Ma received a Ph.D in Electrical and Computer Engineering from The University of Texas at Austin. Contact him at xiaoyuma@google.com.

DAVID PATTERSON is a Distinguished Engineer at Google DeepMind, Mountain View, California, 94043; Board Chair of Laude Institute; and a Pardee professor emeritus at University of California, Berkeley. His research interests include domain-specific computer architectures, AI's environmental impact, and helping shape AI for the public good. Patterson received a Ph.D in Computer Science from University of California, Los Angeles. Contact him at pattrsn@cs.berkeley.edu.

# APPENDIX: DDR DRAM PRICE 1957-2024

See subsequent pages attached. This data is from https://jcmit.net/memoryprice, collected by John C. McCallum. This web site is no longer available. McCallum's last posting was that he was getting older and needed to pass this data collection on to the next generation.

Appendix for "Challenges and Research Directions for Large Language Model Inference Hardware" found in IEEE Computer. This data is from https://jcmit.net/memoryprice, collected by John C. McCallum. This web site is no longer available. McCallum's last posting was that he was gettnig older and needed to pass this data colleciton on to the next generation.

| Memory Price History | | | C:/My Documents/Research/Memprice/memory.xls | | | | | | | copyright 2002, John C. McCallum | | |
|---|---|---|---|---|---|---|---|---|---|---|---|---|
| X | Y | Date | | Ref: | Page | Company | Size | Cost | Speed | Memory Type | JDR Chip Prices | |
| date | $/Gbyte | | | | | | KByte | US $ | nsec | | Size | US$ |
| | | | | | | | | | | | Kbit | |
| 1957 | $411,041,792,000 | 1957 | | Phister 366 | 366 | C.C.C. | 0.00098 | 392 | 10000 | Flip-Flop transistor | | |
| 1959 | $67,947,724,800 | 1959 | | Phister 366 | 366 | E.E.Co. | 0.00098 | 64.8 | 10000 | Flip-Flop - vacuum tube | | |
| 1960 | $5,242,880,000 | 1960 | | Phister 367 | 367 | IBM | 0.00098 | 5 | 11500 | Core memory for IBM 1401 | | |
| 1965 | $2,642,411,520 | 1965 | | Phister 367 | 367 | IBM | 0.00098 | 2.52 | 2000 | Core Memory for IBM 360/30 | | |
| 1970 | $734,003,200 | 1970 | | Phister 367 | 367 | IBM | 0.00098 | 0.7 | 770 | Core Memory for IBM 370/135 | | |
| 1973 | $399,360,000 | 1973 | Jan | PDP8/e User Price List | | DEC | 12 | 4680 | | Core memory 8Kwords x 12 bit | | |
| 1974 | $314,572,800 | 1974 | | Phister 367 | 367 | IBM | 0.00098 | 0.3 | 800 | IC Memory for IBM 370/125 | | |
| 1975 | $421,888,000 | 1975 | Jan | Radio-Electronics | | MITS | 0.25 | 103 | 1000 | Altair 8800 256 Byte Static Board | | |
| 1975.08 | $180,224,000 | 1975 | Feb | | | MITS | 1 | 176 | | Altair 1K Static Board | | |
| 1975.25 | $67,584,000 | 1975 | Apr | | | MITS | 4 | 264 | | Altair 4K DRAM Board | | |
| 1975.75 | $49,920,000 | 1975 | Oct | | | MITS | 4 | 195 | | Altair 4K Static (2102) RAM Board | | |
| 1976 | $40,704,000 | 1976 | Jan | | | MITS | 4 | 159 | | Altair 4K Static (2102) RAM Board | | |
| 1976.17 | $48,960,000 | 1976 | Mar | | | MITS | 16 | 765 | | Altair 16K Static RAM Board | | |
| 1976.42 | $23,040,000 | 1976 | Jun | | | SD Sales | 4 | 90 | | SD Sales 4K Static Board | | |
| 1976.58 | $32,000,000 | 1976 | Aug | | | | 8 | 250 | | 8K Static RAM Board | | |
| 1977.08 | $36,800,000 | 1977 | Feb | | | TDL | 16 | 575 | | S-100 16K | | |
| 1978.17 | $28,000,000 | 1978 | Mar | | | | 64 | 1750 | | S-100 64K | | |
| 1978.25 | $29,440,000 | 1978 | Apr | | | | 16 | 460 | | | | |
| 1978.33 | $19,200,000 | 1978 | May | | | | 16 | 300 | | | | |
| 1978.5 | $24,000,000 | 1978 | Jul | | | Extensis | 64 | 1500 | | | | |
| 1978.58 | $16,000,000 | 1978 | Aug | | | | 8 | 125 | | | | |
| 1978.75 | $15,200,000 | 1978 | Oct | | | | 32 | 475 | | | | |
| 1979 | $10,528,000 | 1979 | Jan | Interface Ag | 124 | | 32 | 329 | | | | |
| 1979.75 | $6,704,000 | 1979 | Oct | | | SD Sales - J | 64 | 419 | | S-100, SD Sales/Jade 64K Kit | | |
| 1980 | $6,480,000 | 1980 | Jan | Interface Ag | 121 | | 64 | 405 | | | | |
| 1981 | $8,800,000 | 1981 | Jan | Interface Ag | 141 | | 64 | 550 | | | | |
| 1981.58 | $4,479,200 | 1981 | Aug | | | Jade | 64 | 279.95 | | | | |
| 1982 | $3,520,000 | 1982 | Jan | Interface Ag | 135 | | 256 | 880 | | | | |
| 1982.17 | $4,464,000 | 1982 | Mar | Microsystems | | | 64 | 279 | | | | |
| 1982.67 | $1,980,000 | 1982 | Sep | BYTE | | California D | 256 | 495 | | | | |
| 1983 | $2,396,000 | 1983 | Jan | Interface Ag | 153 | | 256 | 599 | | | | |
| 1983.67 | $1,980,000 | 1983 | Sep | BYTE | | California D | 256 | 495 | | | | |
| 1984 | $1,378,667 | 1984 | Jan | BYTE | 64 | | 384 | 517 | | | | |
| 1984.58 | $1,330,667 | 1984 | Aug | BYTE | 467 | Advanced Com | 384 | 499 | | IBM PC Board, 384K, $199+6*50 | 256 | 79 |
| 1985 | $880,000 | 1985 | Jan | BYTE | 470 | Do Kay | 512 | 440 | | IBM 512K $199.95+8*29.97 | | |
| 1985.33 | $720,000 | 1985 | May | BYTE | 507 | Do Kay | 512 | 360 | | IBM 512K $199.95+8*19.98 | | 8.95 |
| 1985.42 | $550,000 | 1985 | Jun | BYTE | 505 | Do Kay | 512 | 275 | | IBM 512K $149+8*15.75 | | 5.95 |
| 1985.5 | $420,000 | 1985 | Jul | BYTE | 435 | Fortron | 512 | 210 | | IBM 512K $119+7*13 | | 5.95 |
| 1985.58 | $349,500 | 1985 | Aug | BYTE | 418 | Jade | 2048 | 699 | | 2MB J-RAM-2 A&T $699 | 256 | 3.95 |
| 1985.67 | $299,500 | 1985 | Sep | BYTE | 444 | Jade | 2048 | 599 | | 2MB J-RAM-2 A&T $599 | 256 | 2.95 |
| 1985.83 | $299,500 | 1985 | Nov | BYTE | | Jade | 2048 | 599 | | 2MB J-RAM-2 A&T $599 | 256 | 2.95 |
| 1985.92 | $299,500 | 1985 | Dec | BYTE | | Jade | 2048 | 599 | | 2MB J-RAM-2 A&T $599 | 256 | 2.95 |
| 1986 | $299,500 | 1986 | Jan | BYTE | | Jade | 2048 | 599 | | 2MB J-RAM-2 A&T $599 | 256 | 2.95 |
| 1986.08 | $299,500 | 1986 | Feb | BYTE | | Jade | 2048 | 599 | | 2MB J-RAM-2 A&T $599 | 256 | 2.95 |
| 1986.17 | $299,500 | 1986 | Mar | BYTE | | Jade | 2048 | 599 | | 2MB J-RAM-2 A&T $599 | 256 | 2.95 |
| 1986.25 | $299,500 | 1986 | Apr | BYTE | | Jade | 2048 | 599 | | 2MB J-RAM-2 A&T $599 | 256 | 2.95 |
| 1986.33 | $189,500 | 1986 | May | BYTE | | JDR | 3072 | 568.5 | | JDR AT Multifuntion $199.95+49.95+318.60 | 256 | 2.95 |
| 1986.42 | $189,500 | 1986 | Jun | BYTE | | JDR | 3072 | 568.5 | | JDR AT Multifuntion $199.95+49.95+318.60 | 256 | 2.95 |
| 1986.5 | $189,500 | 1986 | Jul | BYTE | | JDR | 3072 | 568.5 | | JDR AT Multifuntion $199.95+49.95+318.60 | 256 | 2.95 |
| 1986.58 | $189,500 | 1986 | Aug | BYTE | | JDR | 3072 | 568.5 | | JDR AT Multifuntion $199.95+49.95+318.60 | 256 | 2.95 |
| 1986.67 | $189,500 | 1986 | Sep | BYTE | | JDR | 3072 | 568.5 | | JDR AT Multifuntion $199.95+49.95+318.60 | 256 | 2.95 |
| 1986.75 | $189,500 | 1986 | Oct | BYTE | | JDR | 3072 | 568.5 | | JDR AT Multifuntion $199.95+49.95+318.60 | 256 | 2.95 |
| 1986.92 | $189,500 | 1986 | Dec | BYTE | | JDR | 3072 | 568.5 | | JDR AT Multifuntion $199.95+49.95+318.60 | 256 | 2.95 |
| 1987 | $176,167 | 1987 | Jan | BYTE | 477 | JDR | 3072 | 528.5 | | JDR AT Multifuntion $139.95+49.95+3*36*2.95 | 256 | 2.95 |
| 1987.08 | $176,167 | 1987 | Feb | BYTE | | JDR | 3072 | 528.5 | | JDR AT Multifuntion $139.95+49.95+3*36*2.95 | 256 | 2.95 |

| 1987.17 | $157,000 | 1987 | Mar | BYTE | 383 | Pine | 3072 | 471 | | Pine 3MB Multifunction $147+108*3 | 256 | 2.95 |
|---|---|---|---|---|---|---|---|---|---|---|---|---|
| 1987.25 | $153,675 | 1987 | Apr | BYTE | | JDR | 4096 | 614.7 | | JDR MCT-ATRAM 4MB $149.95+39.95+4*36*2.95 | 256 | 2.95 |
| 1987.33 | $153,675 | 1987 | May | BYTE | | JDR | 4096 | 614.7 | | JDR MCT-ATRAM 4MB $149.95+39.95+4*36*2.95 | 256 | 2.95 |
| 1987.42 | $153,675 | 1987 | Jun | BYTE | | JDR | 4096 | 614.7 | | JDR MCT-ATRAM 4MB $149.95+39.95+4*36*2.95 | 256 | 2.95 |
| 1987.5 | $153,675 | 1987 | Jul | BYTE | | JDR | 4096 | 614.7 | | JDR MCT-ATRAM 4MB $149.95+39.95+4*36*2.95 | 256 | 2.95 |
| 1987.58 | $153,675 | 1987 | Aug | BYTE | | JDR | 4096 | 614.7 | | JDR MCT-ATRAM 4MB $149.95+39.95+4*36*2.95 | 256 | 2.95 |
| 1987.67 | $162,823 | 1987 | Sep | BYTE | | JDR | 3072 | 488.47 | | JDR MCT-ATMF 3MB Multifunction | 256 | 2.95 |
| 1987.75 | $133,000 | 1987 | Oct | BYTE | 322 | Advanced Com | 3072 | 399 | | AST Advantage AT w/3MB | 256 | 2.95 |
| 1987.83 | $162,823 | 1987 | Nov | BYTE | | JDR | 3072 | 488.47 | | JDR MCT-ATMF 3MB Multifunction | 256 | 2.95 |
| 1987.92 | $162,823 | 1987 | Dec | BYTE | | JDR | 3072 | 488.47 | | JDR MCT-ATMF 3MB Multifunction | 256 | 2.95 |
| 1988 | $162,823 | 1988 | Jan | BYTE | | JDR | 3072 | 488.47 | | JDR MCT-ATMF 3MB Multifunction | 256 | 2.95 |
| 1988.08 | $182,273 | 1988 | Feb | BYTE | | JDR | 3072 | 546.82 | | JDR MCT-ATMF 3MB Multifunc $169.90+3*36*3.49 | 256 | 3.49 |
| 1988.17 | $198,833 | 1988 | Mar | BYTE | | JDR | 3072 | 596.5 | | JDR MCT-ATMF 3MB Multifunc $169.90+426.60 | 256 | 3.95 |
| 1988.33 | $198,833 | 1988 | May | BYTE | | JDR | 3072 | 596.5 | | JDR MCT-ATMF 3MB Multifunc $169.90+426.60 | 256 | 3.95 |
| 1988.42 | $198,833 | 1988 | Jun | BYTE | | JDR | 3072 | 596.5 | | JDR MCT-ATMF 3MB Multifunc $169.90+426.60 | 256 | 3.95 |
| 1988.5 | $504,833 | 1988 | Jul | BYTE | | JDR | 3072 | 1514.5 | | JDR MCT-ATMF 3MB Multifunc $169.90+3*36*12.45 | 256 | 12.45 |
| 1988.58 | $504,833 | 1988 | Aug | BYTE | | JDR | 3072 | 1514.5 | | JDR MCT-ATMF 3MB Multifunc $169.90+3*36*12.45 | 256 | 12.45 |
| 1988.67 | $504,833 | 1988 | Sep | BYTE | | JDR | 3072 | 1514.5 | | JDR MCT-ATMF 3MB Multifunc $169.90+3*36*12.45 | 256 | 12.45 |
| 1988.75 | $504,833 | 1988 | Oct | BYTE | | JDR | 3072 | 1514.5 | | JDR MCT-ATMF 3MB Multifunc $169.90+3*36*12.45 | 256 | 12.45 |
| 1988.83 | $504,833 | 1988 | Nov | BYTE | | JDR | 3072 | 1514.5 | | JDR MCT-ATMF 3MB Multifunc $169.90+3*36*12.45 | 256 | 12.45 |
| 1988.92 | $504,833 | 1988 | Dec | BYTE | | JDR | 3072 | 1514.5 | | JDR MCT-ATMF 3MB Multifunc $169.90+3*36*12.45 | 256 | 12.45 |
| 1989 | $504,833 | 1989 | Jan | BYTE | | JDR | 3072 | 1514.5 | | JDR MCT-ATMF 3MB Multifunc $169.90+3*36*12.45 | 256 | 12.45 |
| 1989.08 | $504,833 | 1989 | Feb | BYTE | | JDR | 3072 | 1514.5 | | JDR MCT-ATMF 3MB Multifunc $169.90+3*36*12.45 | 256 | 12.45 |
| 1989.17 | $504,833 | 1989 | Mar | BYTE | | JDR | 3072 | 1514.5 | | JDR MCT-ATMF 3MB Multifunc $169.90+3*36*12.45 | 256 | 12.45 |
| 1989.25 | $504,833 | 1989 | Apr | BYTE | | JDR | 3072 | 1514.5 | | JDR MCT-ATMF 3MB Multifunc $169.90+3*36*12.45 | 256 | 12.45 |
| 1989.42 | $344,273 | 1989 | Jun | BYTE | | JDR | 3072 | 1032.82 | | JDR MCT-ATMF 3MB Multifunc $169.90+3*36*7.99 | 1024 | 24.95 |
| 1989.5 | $197,250 | 1989 | Jul | BYTE | 316 | Unitex | 4096 | 789 | | Unitex RAMII-EMS - 4MB $249+36*15.00 | 1024 | 19.95 |
| 1989.58 | $188,250 | 1989 | Aug | BYTE | 302 | Unitex | 4096 | 753 | | Everex RAM II 4M-EMS $249+36*14. | 1024 | 19.95 |
| 1989.67 | $188,250 | 1989 | Sep | BYTE | | Unitex | 4096 | 753 | | Everex RAM II 4M-EMS $249+36*14. | 1024 | 13.95 |
| 1989.75 | $127,875 | 1989 | Oct | BYTE | 340 | Unitex | 8192 | 1023 | | Bocaram AT 8MB - $159+72*12 | 1024 | 13.95 |
| 1989.83 | $116,900 | 1989 | Nov | BYTE | | Unitex | 10240 | 1169 | | Everex RAM 10,000 10MB $179+90*11. | 1024 | 13.95 |
| 1989.92 | $113,125 | 1989 | Dec | BYTE | | Unitex | 8192 | 905 | | Bocaram AT 8MB - $149+72*10.50 | 1024 | 11.95 |
| 1990 | $106,375 | 1990 | Jan | BYTE | | Unitex | 8192 | 851 | | Bocaram AT 8MB - $149+72*9.75 | 1024 | 11.95 |
| 1990.17 | $98,250 | 1990 | Mar | BYTE | | Unitex | 8192 | 786 | | Bocaram AT 8MB - $149+72*8.50 | 1024 | 11.95 |
| 1990.33 | $98,250 | 1990 | May | BYTE | | Unitex | 8192 | 786 | | Bocaram AT 8MB - $149+72*8.50 | 1024 | 11.95 |
| 1990.42 | $89,500 | 1990 | Jun | BYTE | | Unitex | 8192 | 716 | | Bocaram AT 8MB - $140+72*8.00 | 1024 | 11.95 |
| 1990.5 | $82,750 | 1990 | Jul | BYTE | | Unitex | 8192 | 662 | | Bocaram AT 8MB - $140+72*7.25 | 1024 | 11.95 |
| 1990.58 | $81,125 | 1990 | Aug | BYTE | | Nevada | 8192 | 649 | | Bocaram AT Plus 8MB - $649 | 1024 | 11.95 |
| 1990.67 | $71,500 | 1990 | Sep | BYTE | 488 | | 8192 | 572 | | Bocaram AT Plus 8MB - $140+72*6.00 | 1024 | 11.95 |
| 1990.75 | $59,000 | 1990 | Oct | BYTE | 343 | IC Express | 1024 | 59 | 80 | 1Mx9-80 SIMM @ $59 | 1024 | 11.95 |
| 1990.83 | $51,000 | 1990 | Nov | BYTE | 453 | IC Express | 1024 | 51 | 80 | 1Mx9-80 SIMM @ $51 | 1024 | 11.95 |
| 1990.92 | $45,500 | 1990 | Dec | BYTE | 388 | IC Express | 1024 | 45.5 | 80 | 1Mx9-80 SIMM @ $45.50 | 1024 | 7.95 |
| 1991 | $44,500 | 1991 | Jan | BYTE | 393 | IC Express | 1024 | 44.5 | 80 | 1Mx9-80 SIMM @ $44.50 | 1024 | 7.95 |
| 1991.08 | $44,500 | 1991 | Feb | BYTE | 335 | IC Express | 1024 | 44.5 | 80 | 1Mx9-80 SIMM @ $44.50 | 1024 | 7.95 |
| 1991.17 | $45,000 | 1991 | Mar | BYTE | 374 | AMT Internat | 1024 | 45 | 100 | 1Mx9-100 SIMM @ $45 | 1024 | 7.95 |
| 1991.25 | $45,000 | 1991 | Apr | BYTE | 373 | AMT Internat | 1024 | 45 | 100 | 1Mx9-100 SIMM @ $45 | 1024 | 7.95 |
| 1991.33 | $45,000 | 1991 | May | BYTE | 356 | AMT Internat | 1024 | 45 | 100 | 1Mx9-100 SIMM @ $45 | 1024 | 7.95 |
| 1991.42 | $43,750 | 1991 | Jun | BYTE | 398 | IC Express | 4096 | 175 | 80 | 4Mx9-80 SIMM @ $175 | 1024 | 6.89 |
| 1991.5 | $43,750 | 1991 | Jul | BYTE | 333 | IC Express | 4096 | 175 | 80 | 4Mx9-80 SIMM @ $175 | 1024 | 6.89 |
| 1991.58 | $41,250 | 1991 | Aug | BYTE | 321 | IC Express | 4096 | 165 | 80 | 4Mx9-80 SIMM @ $165 | 1024 | 6.89 |
| 1991.67 | $46,250 | 1991 | Sep | BYTE | 375 | Microprocess | 4096 | 185 | 80 | 4Mx9-80 SIMM @ $185 | 1024 | 6.95 |
| 1991.75 | $45,000 | 1991 | Oct | BYTE | 306 | Microprocess | 4096 | 180 | 80 | 4Mx9-80 SIMM @ $180 | 1024 | 6.95 |
| 1991.83 | $39,750 | 1991 | Nov | BYTE | 401 | Nevada | 4096 | 159 | 80 | 4Mx9-80 SIMM @ $159 | 1024 | 5.49 |
| 1991.92 | $39,750 | 1991 | Dec | BYTE | 316 | Nevada | 4096 | 159 | 80 | 4Mx9-80 SIMM @ $159 | 1024 | 5.49 |
| 1992 | $36,250 | 1992 | Jan | BYTE | 380 | AMT Internat | 4096 | 145 | 80 | 4Mx9-80 SIMM @ $145 | 1024 | 5.49 |
| 1992.08 | $36,250 | 1992 | Feb | BYTE | 316 | AMT Internat | 4096 | 145 | 80 | 4Mx9-80 SIMM @ $145 | 1024 | 5.49 |
| 1992.17 | $36,250 | 1992 | Mar | BYTE | 332 | AMT Internat | 4096 | 145 | 80 | 4Mx9-80 SIMM @ $145 | 36864 | 179.95 |
| 1992.25 | $34,750 | 1992 | Apr | BYTE | 318 | Sii Micros | 4096 | 139 | | 4Mx9-?? SIMM @ $139 | 36864 | 179.95 |
| 1992.33 | $30,000 | 1992 | May | BYTE | 342 | AMT Internat | 4096 | 120 | 80 | 4Mx9-80 SIMM @ $120 | 36864 | 179.95 |
| 1992.42 | $32,500 | 1992 | Jun | BYTE | 406 | AmRam | 4096 | 130 | 80 | 4Mx9-80 SIMM @ $130 | 36864 | 179.95 |
| 1992.5 | $33,500 | 1992 | Jul | BYTE | 351 | AmRam | 4096 | 134 | 80 | 4Mx9-80 SIMM @ $134 | 36864 | 179.95 |

| | | | | | | | | | | | | | |
|---|---|---|---|---|---|---|---|---|---|---|---|---|---|
| 1992.58 | $31,000 | 1992 | Aug | BYTE | 324 | Worldwide | 4096 | 124 | 70 | 4Mx9-70 SIMM @ $124 | | 9216 | 39.95 |
| 1992.67 | $27,500 | 1992 | Sep | BYTE | 350 | AMT Internat | 4096 | 110 | 80 | 4Mx9-80 SIMM @ $110 | | 9216 | 39.95 |
| 1992.75 | $26,250 | 1992 | Oct | BYTE | 321 | AMT Internat | 4096 | 105 | 80 | 4Mx9-80 SIMM @ $105 | | 36864 | 149.95 |
| 1992.83 | $26,250 | 1992 | Nov | BYTE | 346 | AMT Internat | 4096 | 105 | 80 | 4Mx9-80 SIMM @ $105 | | 36864 | 149.95 |
| 1992.92 | $26,250 | 1992 | Dec | BYTE | 302 | AMT Internat | 4096 | 105 | 80 | 4Mx9-80 SIMM @ $105 | | 36864 | 149.95 |
| 1993 | $33,063 | 1993 | Jan | BYTE | 307 | Cititronics | 16384 | 529 | | 4Mx36-?? SIMM @ $526 | | 36864 | 149.95 |
| 1993.08 | $27,500 | 1993 | Feb | BYTE | 269 | Memory Super | 4096 | 110 | 70 | 4Mx9-70 SIMM @ $110 | | 36864 | 149.95 |
| 1993.17 | $27,500 | 1993 | Mar | BYTE | 239 | Memory Super | 4096 | 110 | 70 | 4Mx9-70 SIMM @ $110 | | 36864 | 149.95 |
| 1993.25 | $27,500 | 1993 | Apr | BYTE | 245 | Memory Super | 4096 | 110 | 70 | 4Mx9-70 SIMM @ $110 | | 36864 | 149.95 |
| 1993.33 | $27,500 | 1993 | May | BYTE | 263 | Memory Super | 4096 | 110 | 70 | 4Mx9-70 SIMM @ $110 | | 36864 | 149.95 |
| 1993.42 | $30,000 | 1993 | Jun | BYTE | 249 | AMT Internat | 1024 | 30 | 100 | 1Mx9-100 SIMM @ $30 | | 36864 | 149.95 |
| 1993.5 | $30,000 | 1993 | Jul | BYTE | 266 | AMT Internat | 1024 | 30 | 100 | 1Mx9-100 SIMM @ $30 | | 36864 | 149.95 |
| 1993.58 | $30,000 | 1993 | Aug | BYTE | 245 | AMT Internat | 1024 | 30 | 100 | 1Mx9-100 SIMM @ $30 | | 36864 | 149.95 |
| 1993.67 | $30,000 | 1993 | Sep | BYTE | 267 | AMT Internat | 1024 | 30 | 100 | 1Mx9-100 SIMM @ $30 | | 36864 | 149.95 |
| 1993.75 | $36,000 | 1993 | Oct | BYTE | 260 | LA Trade | 4096 | 144 | 80 | 4Mx9-80 SIMM @ $144 | | 36864 | 159.95 |
| 1993.83 | $39,750 | 1993 | Nov | BYTE | 360 | First Source | 4096 | 159 | 70 | 4Mx9-70 SIMM @ $159 | | 36864 | 159.95 |
| 1993.92 | $35,750 | 1993 | Dec | BYTE | 280 | West Coast M | 4096 | 143 | 70 | 4Mx3x3-70 SIMM @ $143 | | 36864 | 159.95 |
| 1994 | $35,750 | 1994 | Jan | BYTE | 290 | West Coast M | 4096 | 143 | 70 | 4Mx3x3-70 SIMM @ $143 | | 36864 | 159.95 |
| 1994.08 | $35,750 | 1994 | Feb | BYTE | 250 | West Coast M | 4096 | 143 | 70 | 4Mx3x3-70 SIMM @ $143 | | 36864 | 159.95 |
| 1994.17 | $36,000 | 1994 | Mar | BYTE | 251 | Nevada | 4096 | 144 | 80 | 4Mx9-80 SIMM @ $144 | | 36864 | 159.95 |
| 1994.25 | $37,250 | 1994 | Apr | BYTE | 270 | La Trade | 4096 | 149 | 80 | 4Mx9-80 SIMM @ $149 | | 36864 | 154.95 |
| 1994.33 | $37,250 | 1994 | May | BYTE | 238 | La Trade | 4096 | 149 | 80 | 4Mx9-80 SIMM @ $149 | | 36864 | 154.95 |
| 1994.42 | $37,250 | 1994 | Jun | BYTE | 320 | La Trade | 4096 | 149 | 80 | 4Mx9-80 SIMM @ $149 | | 36864 | 169.95 |
| 1994.5 | $38,500 | 1994 | Jul | BYTE | 239 | Nevada | 4096 | 154 | 80 | 4Mx9-80 SIMM @ $154 | | 36864 | 169.95 |
| 1994.58 | $37,000 | 1994 | Aug | BYTE | 226 | Nevada | 1024 | 37 | 100 | 1Mx9-100 SIMM @ $37 | | 36864 | 169.95 |
| 1994.67 | $34,000 | 1994 | Sep | BYTE | 244 | Pacific Coas | 4096 | 136 | 70 | 4Mx9-70 SIMM @ $136 | | 36864 | 169.95 |
| 1994.75 | $33,500 | 1994 | Oct | BYTE | 259 | Pacific Coas | 4096 | 134 | 70 | 4Mx9-70 SIMM @ $134 | | 36864 | 159.95 |
| 1994.83 | $32,250 | 1994 | Nov | BYTE | 324 | Pacific Coas | 4096 | 129 | 70 | 4Mx9-70 SIMM @ $129 | | 36864 | 159.95 |
| 1994.92 | $32,250 | 1994 | Dec | BYTE | 264 | Pacific Coas | 4096 | 129 | 70 | 4Mx9-70 SIMM @ $129 | | 36864 | 159.95 |
| 1995 | $32,250 | 1995 | Jan | BYTE | 256 | Nevada | 4096 | 129 | 80 | 4Mx9-80 SIMM @ $129 | | 36864 | 159.95 |
| 1995.08 | $32,000 | 1995 | Feb | BYTE | 202 | First Source | 4096 | 128 | 70 | 4Mx9-70 SIMM @ $129 | | 36864 | 159.95 |
| 1995.17 | $32,000 | 1995 | Mar | BYTE | 212 | First Source | 4096 | 128 | 70 | 4Mx9-70 SIMM @ $129 | | 36864 | 159.95 |
| 1995.25 | $31,188 | 1995 | Apr | BYTE | 254 | Pacific Coas | 16384 | 499 | 70 | 4Mx36-70 SIMM [72 pin] @ $499 | | 9216 | 37.95 |
| 1995.33 | $31,188 | 1995 | May | BYTE | 222 | Pacific Coas | 16384 | 499 | 70 | 4Mx36-70 SIMM [72 pin] @ $499 | | 9216 | 37.95 |
| 1995.42 | $31,125 | 1995 | Jun | BYTE | 286 | First Source | 16384 | 498 | 70 | 4Mx36-70 SIMM [72 pin] @ $498 | | 36864 | 135.29 |
| 1995.5 | $31,188 | 1995 | Jul | BYTE | 213 | Pacific Coas | 16384 | 499 | 70 | 4Mx36-70 SIMM [72 pin] @ $499 | | 147456 | 619 |
| 1995.58 | $30,563 | 1995 | Aug | BYTE | 190 | Pacific Coas | 16384 | 489 | 70 | 4Mx36-70 SIMM [72 pin] @ $489 | | 147456 | 619 |
| 1995.67 | $33,063 | 1995 | Sep | BYTE | 306 | Future Micro | 16384 | 529 | 70 | 4Mx36-70 SIMM [72 pin] @ $529 | | 147456 | 619 |
| 1995.75 | $33,063 | 1995 | Oct | BYTE | 214 | Future Micro | 16384 | 529 | 70 | 4Mx36-70 SIMM [72 pin] @ $529 | | 294912 | 1199 |
| 1995.83 | $30,875 | 1995 | Nov | BYTE | 286 | First Source | 16384 | 494 | 70 | 4Mx36-70 SIMM [72 pin] @ $494 | | 294912 | 1199 |
| 1995.92 | $30,875 | 1995 | Dec | BYTE | 244 | First Source | 16384 | 494 | 70 | 4Mx36-70 SIMM [72 pin] @ $494 | | 147456 | 699 |
| 1996 | $29,875 | 1996 | Jan | BYTE | 186 | First Source | 16384 | 478 | 70 | 4Mx36-70 SIMM [72 pin] @ $478 | | 147456 | 689 |
| 1996.08 | $28,750 | 1996 | Feb | BYTE | 217 | Worldwide | 16384 | 460 | 70 | 4Mx36-70 SIMM [72 pin] @ $460 | | 147456 | 689 |
| 1996.17 | $26,125 | 1996 | Mar | BYTE | 170 | First Source | 8192 | 209 | 70 | 2Mx36-70 SIMM [72 pin] @ $209 | | 36864 | 133.95 |
| 1996.25 | $24,688 | 1996 | Apr | BYTE | 196 | Worldwide | 16384 | 395 | 70 | 4Mx36-70 SIMM [72 pin] @ $395 | | 36864 | 129.95 |
| 1996.33 | $17,188 | 1996 | May | BYTE | 195 | Worldwide | 32768 | 550 | 60 | 8Mx36-60 SIMM [72 pin] @ $550 | no JDR ads anymore | | |
| 1996.42 | $14,875 | 1996 | Jun | BYTE | 186 | First Source | 8192 | 119 | 70 | 2Mx36-70 SIMM [72 pin] @ $119 | Byte appears to be | | |
| 1996.5 | $11,250 | 1996 | Jul | BYTE | 180 | Worldwide | 16384 | 180 | 60 | 4Mx32-60 SIMM [72 pin] @ $180 | going downhill in | | |
| 1996.58 | $9,063 | 1996 | Aug | BYTE | 164 | Worldwide | 16384 | 145 | 60 | 4Mx36-60 SIMM [72 pin] @ $145 | subscriptions, etc. | | |
| 1996.67 | $8,438 | 1996 | Sep | BYTE | 192 | Worldwide | 16384 | 135 | 60 | 4Mx36-60 SIMM [72 pin] @ $135 | | | |
| 1996.75 | $8,000 | 1996 | Oct | BYTE | 181 | First Source | 16384 | 128 | 70 | 4Mx36-70 SIMM [72 pin] @ $128 | | | |
| 1996.83 | $5,250 | 1996 | Nov | BYTE | 210 | First Source | 8192 | 42 | 70 | 2Mx32-70 SIMM [72 pin] @ $42 | | | |
| 1996.92 | $5,250 | 1996 | Dec | BYTE | 177 | First Source | 8192 | 42 | 70 | 2Mx32-70 SIMM [72 pin] @ $42 | | | |
| 1997 | $4,625 | 1997 | Jan | BYTE | 153 | First Source | 8192 | 37 | 60 | 2Mx32-60 SIMM EDO [72 pin] @ $37 | | | |
| 1997.08 | $3,625 | 1997 | Feb | BYTE | 169 | Memory On-Li | 8192 | 29 | 60 | 2Mx32-60 SIMM EDO [72 pin] @ $29 | | | |
| 1997.17 | $3,000 | 1997 | Mar | BYTE | 167 | Memory On-Li | 8192 | 24 | 60 | 2Mx32-60 SIMM EDO [72 pin] @ $24 | | | |
| 1997.25 | $3,000 | 1997 | Apr | BYTE | 166 | Memory On-Li | 8192 | 24 | 60 | 2Mx32-60 SIMM EDO [72 pin] @ $24 | | | |
| 1997.33 | $3,000 | 1997 | May | BYTE | 147 | Memory On-Li | 8192 | 24 | 60 | 2Mx32-60 SIMM EDO [72 pin] @ $24 | | | |
| 1997.42 | $3,688 | 1997 | Jun | BYTE | 163 | Memory On-Li | 16384 | 59 | 60 | 4Mx32-60 SIMM EDO [72 pin] @ $59 | | | |
| 1997.5 | $4,000 | 1997 | Jul | Pcmag | 401 | Miami Int'l | 8192 | 32 | 70 | 2Mx32-? SIMM EDO [72 pin] @ $32 | last Byte | | |
| 1997.58 | $4,125 | 1997 | Aug | Pcmag | 416 | Bason Memory | 8192 | 33 | 70 | 2Mx32-? SIMM FPM [72 pin] @ $33 | | | |

| | | | | | | | | | | | |
|---|---|---|---|---|---|---|---|---|---|---|---|
| 1997.67 | $3,625 | 1997 | Sep23 | Pcmag | 306 | Miami Int'l | 16384 | 58 | | ??? | |
| 1997.75 | $3,406 | 1997 | Oct21 | Pcmag | 339 | Miami Int'l | 32768 | 109 | | ??? | |
| 1997.83 | $3,250 | 1997 | Nov18 | Pcmag | 315 | LA Trade | 32768 | 104 | | 8Mx32-? SIMM EDO [72 pin] @ $104 | |
| 1997.92 | $2,156 | 1997 | Dec16 | Pcmag | 329 | Computer Ame | 32768 | 69 | | 8Mx32-? SIMM EDO [72 pin] @ $69 | |
| 1998 | $2,156 | 1998 | Jan20 | Pcmag | 277 | Computer Ame | 32768 | 69 | | 8Mx32-? SIMM EDO [72 pin] @ $69 | |
| 1998.08 | $906 | 1998 | Feb24 | Pcmag | 261 | Computer Ame | 32768 | 29 | | 8Mx32-? SIMM EDO [72 pin] @ $29 | |
| 1998.17 | $969 | 1998 | Mar10 | Pcmag | 302 | Computer Ame | 32768 | 31 | | 8Mx32-? SIMM EDO [72 pin] @ $31 | |
| 1998.25 | $1,219 | 1998 | Apr21 | Pcmag | 258 | Computer Ame | 32768 | 39 | | 8Mx32-? SIMM EDO [72 pin] @ $39 | |
| 1998.33 | $1,188 | 1998 | May26 | Pcmag | 294 | Computer Ame | 32768 | 38 | | 8Mx32-? SIMM EDO [72 pin] @ $38 | |
| 1998.42 | $969 | 1998 | Jun30 | Pcmag | 296 | Computer Ame | 32768 | 31 | | 8Mx32-? SIMM EDO [72 pin] @ $31 | |
| 1998.58 | $1,031 | 1998 | Aug | Pcmag | 388 | Computer Ser | 32768 | 33 | | 4Mx64-? DIMM SDRAM @ $33 | |
| 1998.67 | $969 | 1998 | Sep22 | Pcmag | 279 | Computer Ser | 32768 | 31 | | 8Mx32-? SIMM EDO [72 pin] @ $31 | |
| 1998.75 | $1,156 | 1998 | Oct20 | Pcmag | 293 | Computer Ser | 32768 | 37 | | 8Mx32-? SIMM EDO [72 pin] @ $37 | |
| 1998.83 | $844 | 1998 | Nov17 | Pcmag | 286 | Discount Mem | 32768 | 27 | | 8Mx32-? SIMM FPM [72 pin] @ $27 | |
| 1998.92 | $844 | 1998 | Dec1 | Pcmag | 390 | Discount Mem | 32768 | 27 | | 8Mx32-? SIMM FPM [72 pin] @ $27 | |
| 1999.08 | $1,438 | 1999 | Feb9 | Pcmag | 263 | Memory Liqui | 32768 | 46 | | 8Mx32-? SIMM FPM [72 pin] @ $46 | |
| 1999.13 | $844 | 1999 | Feb23 | Pcmag | 226 | Discount Mem | 32768 | 27 | | 8Mx32-? SIMM FPM [72 pin] @ $27 | |
| 1999.17 | $1,250 | 1999 | Mar23 | Pcmag | 261 | Tiger System | 65536 | 79.99 | | 64 MB DIMM PC-100 @ $79.99 | |
| 1999.25 | $1,250 | 1999 | Apr20 | Pcmag | 277 | TigerDirect. | 65536 | 79.99 | | 64 MB DIMM PC-100 @ $79.99 | |
| 1999.33 | $859 | 1999 | May25 | Pcmag | 273 | TigerDirect. | 65536 | 54.99 | | 64 MB DIMM PC-100 @ $54.99 | |
| 1999.5 | $781 | 1999 | Jul99 | Pcmag | 323 | TigerDirect. | 131072 | 99.99 | | 128 MB DIMM PC-100 @ $99.99 | |
| 1999.67 | $868 | 1999 | Sep21 | Pcmag | 222 | www.crucial. | 131072 | 111.14 | | 128 MB DIMM PC-100 @ $111.14 | |
| 1999.75 | $1,039 | 1999 | Oct19 | Pcmag | 219 | www.crucial. | 65536 | 66.49 | | 64 MB DIMM PC-100 @ $66.49 | |
| 1999.83 | $1,336 | 1999 | Nov16 | Pcmag | 247 | www.crucial. | 131072 | 170.99 | | 128 MB DIMM PC-100 @ $170.99 | |
| 1999.92 | $2,348 | 1999 | Dec1 | Pcmag | 287 | www.crucial. | 131072 | 300.59 | | 128 MB DIMM PC-100 @ $300.59 | |
| 2000 | $1,561 | 2000 | Jan18 | Pcmag | 14 | Crucial Tech | 65536 | 99.89 | | 64 MB DIMM PC-100 @ $99.89 | |
| 2000.08 | $1,476 | 2000 | Feb18 | Pcmag | 18 | Crucial Tech | 65536 | 94.49 | | 64 MB DIMM PC-100 @ $94.49 | |
| 2000.17 | $1,078 | 2000 | Mar21 | Pcmag | 214 | StarSurplus. | 65536 | 69 | | 64 MB DIMM PC-100 @ $69 | |
| 2000.25 | $844 | 2000 | Apr18 | Pcmag | 16 | Crucial Tech | 65536 | 53.99 | | 64 MB DIMM PC-100 @ $53.99 | |
| 2000.33 | $695 | 2000 | May9 | Pcmag | 254 | StarSurplus. | 131072 | 89 | | 128 MB DIMM PC-100 @ $89 | |
| 2000.42 | $900 | 2000 | Jun27 | Pcmag | 18 | Crucial Tech | 65536 | 57.59 | | 64 MB DIMM PC-100 @ $57.59 | |
| 2000.5 | $773 | 2000 | Jul | Pcmag | 228 | StarSurplus. | 131072 | 99 | | 128 MB DIMM PC-133 @ $99 | |
| 2000.58 | $844 | 2000 | Aug | Pcmag | 18 | Crucial Tech | 65536 | 53.99 | | 64 MB DIMM PC-100 @ $53.99 | |
| 2000.67 | $1,069 | 2000 | Sep1 | Pcmag | 18 | Crucial Tech | 65536 | 68.39 | | 64 MB DIMM PC-133 @ $68.39 | only advert for RAM |
| 2000.75 | $1,125 | 2000 | Oct3 | Pcmag | 18 | Crucial Tech | 65536 | 71.99 | | 64 MB DIMM PC-133 @ $71.99 | only advert for RAM |
| 2000.83 | $1,125 | 2000 | Nov7 | Pcmag | 18 | Crucial Tech | 65536 | 71.99 | | 64 MB DIMM PC-133 @ $71.99 | only advert for RAM |
| 2000.92 | $900 | 2000 | Dec5 | Pcmag | 19 | Crucial Tech | 65536 | 57.59 | | 64 MB DIMM PC-133 @ $57.59 | discount for web purchase |
| 2001 | $745 | 2001 | Jan2 | Pcmag | 19 | Crucial Tech | 65536 | 47.69 | | 64 MB DIMM PC-133 @ $47.69 | discount for web purchase |
| 2001.08 | $464 | 2001 | Feb6 | Pcmag | 90 | Crucial Tech | 131072 | 59.39 | | 128 MB DIMM PC-133 @ $59.39 | discount for web purchase |
| 2001.17 | $464 | 2001 | Mar6 | Pcmag | 99 | Crucial Tech | 131072 | 59.39 | | 128 MB DIMM PC-133 @ $59.39 | discount for web purchase |
| 2001.25 | $383 | 2001 | Apr3 | Pcmag | 178 | StarSurplus. | 131072 | 49 | | 128 MB DIMM PC-133 @ $49 | |
| 2001.33 | $387 | 2001 | May8 | Pcmag | 85 | Crucial Tech | 131072 | 49.49 | | 128 MB DIMM PC-100 @ $49.49 | discount for web purchase |
| 2001.42 | $305 | 2001 | Jun12 | Pcmag | 207 | StarSurplus. | 131072 | 39 | | 128 MB DIMM PC-133 @ $39 | |
| 2001.5 | $352 | 2001 | Jul | Pcmag | 143 | Crucial Tech | 262144 | 89.99 | | 256 MB DDR PC2100 @ $89.99 | discount for web purchase |
| 2001.5 | $270 | 2001 | Jul | Pcmag | 196 | iMemoryMall. | 262144 | 69 | | 256 MB DIMM PC-133 @ $69 | |
| 2001.58 | $191 | 2001 | Aug | Pcmag | 169 | StarSurplus. | 262144 | 49 | | 256 MB DIMM PC-133 @ $49 | |
| 2001.67 | $191 | 2001 | Sep4 | Pcmag | 215 | StarSurplus. | 262144 | 49 | | 256 MB DIMM PC-133 @ $49 | |
| 2001.75 | $169 | 2001 | Oct16 | Pcmag | 60 | Crucial Tech | 131072 | 21.59 | | 128 MB DIMM PC-133 @ $21.59 | discount for web purchase |
| 2001.77 | $148 | 2001 | Oct30 | Pcmag | 161 | Crucial Tech | 131072 | 18.89 | | 128 MB DIMM PC-133 @ $18.89 | discount for web purchase |
| 2002.08 | $134 | 2002 | Feb12 | Pcmag | 47 | Crucial Tech | 262144 | 34.19 | | 256 MB DIMM PC-133 @ $34.19 | |
| 2002.08 | $207 | 2002 | Feb12 | Pcmag | 47 | Crucial Tech | 262144 | 53.09 | | 256 MB DDR PC2100 @ $53.09 | |
| 2002.25 | $193 | 2002 | Apr9 | Pcmag | 133 | StarSurplus. | 524288 | 99 | | 512 MB DIMM PC-133 @ $99 | |
| 2002.33 | $193 | 2002 | May7 | Pcmag | 132 | StarSurplus. | 524288 | 99 | | 512 MB DIMM PC-133 @ $99 | |
| 2002.42 | $330 | 2002 | Jun30 | Pcmag | 68 | Crucial Tech | 131072 | 42.29 | | 128 MB DIMM PC-133 @ $42.29 | |
| 2002.58 | $193 | 2002 | Aug02 | Pcmag | 49 | Crucial Tech | 262144 | 49.49 | | 256 MB DIMM PC-133 @ $49.49 | |
| 2002.75 | $193 | 2002 | Oct1 | Pcmag | 115 | Crucial Tech | 262144 | 49.49 | | 256 MB DIMM PC-133 @ $49.49 | |
| 2003.17 | $176 | 2003 | Mar25 | Pcmag | 95 | NewEgg.com | 262144 | 45 | | 256 MB DIMM DDR-2100 @ $45 | |
| 2003.25 | $76 | 2003 | Apr22 | Pcmag | 133 | StarSurplus. | 524288 | 39 | | 512 MB DIMM PC-133 @ $39 | |
| 2003.33 | $126 | 2003 | May27 | Pcmag | 95 | NewEgg.com | 524288 | 64.5 | | 512 MB DIMM DDR-2700 @ $64.50 | |
| 2003.42 | $115 | 2003 | Jun17 | Pcmag | 137 | HardDrive.co | 524288 | 59 | | 512 MB DIMM PC-133 @ $59 | |
| 2003.5 | $133 | 2003 | July | Pcmag | 101 | Crucial Tech | 262144 | 33.99 | | 256 MB DIMM DDR-2100 @ $33.99 | |

| 2003.58 | $129 | 2003 | Aug19 | Pcmag | 54 | Crucial Tech | 524288 | 65.99 | | 512 MB DIMM SDRAM @ $65.99 | DDR at $72.99 |
|---|---|---|---|---|---|---|---|---|---|---|---|
| 2003.67 | $143 | 2003 | Sep16 | Pcmag | 64 | Crucial Tech | 524288 | 72.99 | | 512 MB DIMM SDRAM @ $72.99 | DDR at $86.99 |
| 2003.75 | $148 | 2003 | Oct1 | Pcmag | 54 | Crucial Tech | 524288 | 75.99 | | 512 MB DIMM SDRAM @ $75.99 | DDR at $86.99 |
| 2003.83 | $160 | 2003 | Nov25 | Pcmag | 82 | Crucial Tech | 524288 | 81.99 | | 512 MB DIMM PC-133 @ $81.99 | DDR at $85.99 |
| 2003.99 | $166 | 2003 | Dec30 | Pcmag | 37 | Crucial Tech | 524288 | 84.99 | | 512 MB DIMM PC-133 @ $84.99 | DDR at $85.99 |
| 2004 | $174 | 2004 | Jan | Pcmag | 143 | NewEgg.com | 524288 | 89 | | 512 MB DIMM DDR-3200 @ $89 | |
| 2004.08 | $148 | 2004 | Feb17 | Pcmag | 78 | NewEgg.com | 524288 | 76 | | 512 MB DIMM DDR-3200 @ $76 | GEIL |
| 2004.17 | $146 | 2004 | Mar16 | Pcmag | 115 | NewEgg.com | 524288 | 75 | | 512 MB DIMM DDR-3200 @ $75 | GEIL |
| 2004.33 | $156 | 2004 | May4 | Pcmag | 19 | NewEgg.com | 524288 | 80 | | 512 MB DIMM DDR-3200 @ $76 | GEIL |
| 2004.42 | $203 | 2004 | Jun8 | Pcmag | 19 | NewEgg.com | 524288 | 104 | | 512 MB DIMM DDR-3200 @ $104 | GEIL |
| 2004.5 | $176 | 2004 | July | Pcmag | 19 | NewEgg.com | 524288 | 90 | | 512 MB DIMM DDR-3200 @ $90 | Kingston |
| 2005.25 | $185 | 2005 | Apr12 | Pcmag | 24 | NewEgg.com | 1048576 | 189 | | 2x512 MB DIMM DDR-3200 @ $90 | Mushkin |
| 2005.42 | $149 | 2005 | Jun7 | Pcmag | 20 | NewEgg.com | 1048576 | 153 | | 1 GB DIMM DDR PC-3200Pro @ $153 | Corsair |
| 2005.83 | $116 | 2005 | Nov8 | Pcmag | 67 | NewEgg.com | 1048576 | 119 | | 1 GB DIMM DDR2-533 @ $119 | |
| 2005.92 | $185 | 2005 | Dec6 | Pcmag | 139 | NewEgg.com | 1048576 | 189 | | 2x512 MB DIMM DDR-4800 @ $189 | |
| 2006.17 | $112 | 2006 | Mar21 | Pcmag | 66 | NewEgg.com | 2097152 | 229.81 | | 2x1 GB DIMM DDR-500 PC4000 @ $229.81 | OCZ Gold |
| 2006.33 | $73 | 2006 | May25 | Web | | NewEgg.com | 2097152 | 148.99 | | 2x1 GB DIMM DDR2-667 @ $148.99 free shipping | OCZ Gold |
| 2006.5 | $82 | 2006 | Jul23 | Web | | NewEgg.com | 1048576 | 83.99 | | 1GB DIMM DDR2-667 @$83.99 | Kingston |
| 2006.67 | $73 | 2006 | Sep1 | Web | | NewEgg.com | 2097152 | 149.99 | | 2x1 GB DIMM DDR2-667 @ $179.99-$30 rebate + 4.99 | OCZ Gold |
| 2006.75 | $88 | 2006 | Oct5 | Web | | NewEgg.com | 2097152 | 179.98 | | 2x1 GB DIMM DDR2-667 @ $214.99-$40 rebate + 4.99 | PQi |
| 2006.83 | $98 | 2006 | Nov25 | Web | | NewEgg.com | 2097152 | 199.99 | | 2x1 GB DIMM DDR2-800 @ $199.99 + free shipping | G. Skill |
| 2006.99 | $92 | 2006 | Dec21 | Web | | NewEgg.com | 1048576 | 93.98 | | 1GB DIMM DDR-400 @$88.99 + 4.99 shipping | Wintec |
| 2007 | $82 | 2007 | Jan28 | Web | | NewEgg.com | 1048576 | 83.98 | | 1GB DIMM DDR-400 @$78.99 + 4.99 shipping | PQi |
| 2007.08 | $78 | 2007 | Feb28 | Web | | NewEgg.com | 1048576 | 79.98 | | 1GB DIMM DDR2-800 @ $74.99 + $4.99 shipping | Patriot |
| 2007.17 | $66 | 2007 | Mar15 | Web | | NewEgg.com | 2097152 | 134.98 | | 2x1 GB DIMM DDR2-667 @ $129.99 + 4.99 shipping | Wintec |
| 2007.33 | $46 | 2007 | May7 | Web | | NewEgg.com | 2097152 | 94.99 | 5-5-5-18 | 2x1 GB DIMM DDR2-800 @ $94.99 + free shipping | A-Data |
| 2007.5 | $39 | 2007 | July9 | Web | | NewEgg.com | 2097152 | 78.98 | cas 5 | 2x1 GB DIMM DDR2-667 @ $73.99 + 4.99 shipping | Super Talent |
| 2007.67 | $35 | 2007 | Sept30 | Web | | NewEgg.com | 2097152 | 71.98 | cas 5 | 2x1 GB DIMM DDR2-800 @ $66.99 + 4.99 shipping | Mushkin |
| 2007.75 | $32 | 2007 | Oct16 | Web | | NewEgg.com | 2097152 | 65.98 | 5-5-5-18 | 2x1 GB DIMM DDR2-800 @ $60.99 + 4.99 shipping | A-Data |
| 2007.83 | $24 | 2007 | Nov10 | Web | | NewEgg.com | 2097152 | 49.98 | | 2x1 GB DIMM DDR2-667 @ $44.99 + 4.99 shipping | Kingston |
| 2007.92 | $24 | 2007 | Dec05 | Web | | NewEgg.com | 2097152 | 49.95 | 5-5-5-15 | 2x1 GB DIMM DDR2-800 @ $49.95 + free shipping | G.Skill |
| 2008 | $23 | 2008 | Jan18 | Web | | NewEgg.com | 4194304 | 94.99 | 5-5-5-15 | 2x2 GB DIMM DDR2-800 @ $94.99 + free shipping | G.Skill |
| 2008.08 | $22 | 2008 | Feb20 | Web | | NewEgg.com | 2097152 | 44.99 | 5-5-5-15 | 2x1 GB DIMM DDR2-800 @ $44.99 + free shipping | G.Skill |
| 2008.33 | $22 | 2008 | May18 | Web | | NewEgg.com | 2097152 | 44.99 | 5-5-5-15 | 2x1 GB DIMM DDR2-800 @ $44.99 + free shipping | G.Skill |
| 2008.5 | $21 | 2008 | Jul07 | Web | | NewEgg.com | 4194304 | 84.98 | 5-5-5-18 | 2x2 GB DIMM DDR2-800 @ $77.99 + 5.99 shipping | A-Data |
| 2008.58 | $18 | 2008 | Aug21 | Web | | NewEgg.com | 2097152 | 35.99 | | 2x1 GB DIMM DDR2-667 @ $35.99 + free shipping | Crucial |
| 2008.67 | $15 | 2008 | Sep25 | Web | | NewEgg.com | 2097152 | 29.99 | cas 5 | 2x1 GB DIMM DDR2-667 @ $29.99 + free shipping | Crucial |
| 2008.83 | $11 | 2008 | Nov14 | Web | | NewEgg.com | 4194304 | 44.99 | 5-5-5-15 | 2x2 GB DIMM DDR2-800 @ $49.99 + free shipping | G.Skill |
| 2008.92 | $10 | 2008 | Dec14 | Web | | NewEgg.com | 4194304 | 39.99 | 5-5-5-15 | 2x2 GB DIMM DDR2-800 @ $39.99 + free shipping | G.Skill |
| 2009 | $10 | 2009 | Jan19 | Web | | NewEgg.com | 4194304 | 39.99 | 5-5-5-15 | 2x2 GB DIMM DDR2-800 @ $39.99 + free shipping | G.Skill |
| 2009.08 | $11 | 2009 | Feb14 | Web | | NewEgg.com | 2097152 | 21.99 | cas 6 | 2x1 GB DIMM DDR2-800 @ $21.99 + free shipping | Kingston |
| 2009.25 | $10 | 2009 | Apr03 | Web | | NewEgg.com | 4194304 | 42.99 | 5-5-5-15 | 2x2 GB DIMM DDR2-800 @ $42.99 + free shipping | G.Skill |
| 2009.42 | $11 | 2009 | Jun03 | Web | | NewEgg.com | 4194304 | 46.99 | 5-5-5-18 | 2x2 GB DIMM DDR2-800 @ $46.99 + free shipping | Corsair |
| 2009.5 | $11 | 2009 | Jul02 | Web | | NewEgg.com | 4194304 | 44.99 | 5-5-5-18 | 2x2 GB DIMM DDR2-800 @ $44.99 + free shipping | Corsair |
| 2009.58 | $13 | 2009 | Aug20 | Web | | NewEgg.com | 4194304 | 51.99 | 5-5-5-15 | 2x2 GB DIMM DDR2-800 @ $51.99 + free shipping | G.Skill |
| 2009.75 | $18 | 2009 | Oct07 | Web | | NewEgg.com | 4194304 | 74.99 | 5-5-5-15 | 2x2 GB DIMM DDR2-1066 @ $74.99 + free shipping | Patriot |
| 2009.92 | $21 | 2009 | Dec12 | Web | | NewEgg.com | 2097152 | 41.99 | | 1x2 GB DIMM DDR2-800 @ $41.99 + free shipping | Crucial |
| 2010 | $19 | 2010 | Jan15 | Web | | NewEgg.com | 4194304 | 77.99 | 5-5-5-15 | 2x2 GB DIMM DDR2-800 @ $77.99 + free shipping | Wintec |
| 2010.08 | $20 | 2010 | Feb11 | Web | | NewEgg.com | 4194304 | 82.78 | cas 5 | 2x2 GB DIMM DDR2-667 @ $79.99 + 2.99 shipping | Patriot |
| 2010.17 | $20 | 2010 | Mar10 | Web | | NewEgg.com | 2097152 | 39.99 | cas 5 | 2x1 GB DIMM DDR2-800 @ $39.99 + free shipping | Wintec |
| 2010.33 | $24 | 2010 | May09 | Web | | NewEgg.com | 4194304 | 98.98 | cas 5 | 2x2 GB DIMM DDR2-800 @ $92.99 + 5.99 shipping | PQi |
| 2010.5 | $21 | 2010 | Jul25 | Web | | NewEgg.com | 4194304 | 85.98 | 9-9-9-20 | 2x2 GB DIMM DDR3-1333 @ $84.99 + 0.99 shipping | |
| 2010.58 | $22 | 2010 | Aug26 | Web | | NewEgg.com | 4194304 | 89.99 | 9-9-9-24 | 2x2 GB DIMM DDR3-1333 @ $89.99 + free shipping | Patriot |
| 2010.75 | $17 | 2010 | Oct07 | Web | | NewEgg.com | 4194304 | 69.99 | 7-7-7-20 | 2x2 GB DIMM DDR3-1066 @ $69.99 + free shipping | OCZ Gold |
| 2010.83 | $15 | 2010 | Nov16 | Web | | NewEgg.com | 4194304 | 59.99 | 9-9-9-24 | 2x2 GB DIMM DDR3-1333 @ $59.99 + free shipping | Mushkin |
| 2010.92 | $12 | 2010 | Dec10 | Web | | NewEgg.com | 8388608 | 99.99 | 9-9-9-24 | 2x4 GB DIMM DDR3-1333 @ $99.99 + free shipping | Mushkin |
| 2011 | $10 | 2011 | Jan21 | Web | | NewEgg.com | 4194304 | 40.98 | cas 9 | 2x2 GB DIMM DDR3-1333 @ $40.98 + free shipping | A-Data |
| 2011.08 | $10 | 2011 | Feb12 | Web | | NewEgg.com | 4194304 | 41.99 | cas 9 | 1x4 GB DIMM DDR3-1333 @ $41.99 + free shipping | Kingston |
| 2011.33 | $10 | 2011 | May06 | Web | | NewEgg.com | 8388608 | 81.99 | 9-9-9-24 | 2x4 GB DIMM DDR3-1333 @ $81.99 + free shipping | GEIL |
| 2011.42 | $9 | 2011 | Jun22 | Web | | NewEgg.com | 8388608 | 69.99 | cas 9 | 2x4 GB DIMM DDR3-1333 @ $69.99 + free shipping | Kingston |
| 2011.67 | $5 | 2011 | Sep02 | Web | | NewEgg.com | 4194304 | 21.99 | cas 9 | 1x4 GB DIMM DDR3-1333 @ $21.99 + free shipping | Crucial |

| 2011.75 | $5 | 2011 | Oct17 | Web | | NewEgg.com | 8388608 | 41.99 | cas 9 | 2x4 GB DIMM DDR3-1333 @ $41.99 + free shipping | PNY Optima | |
|---|---|---|---|---|---|---|---|---|---|---|---|---|
| 2012 | $5 | 2012 | Jan12 | Web | | NewEgg.com | 8388608 | 39.99 | 9-9-9-24 | 2x4 GB DIMM DDR3-1600 @ $39.99 + free shipping | Corsair | |
| 2012.08 | $5 | 2012 | Feb13 | Web | | NewEgg.com | 8388608 | 39.99 | cas 9 | 2x4 GB DIMM DDR3-1600 @ $39.99 + free shipping | Kingston | |
| 2012.25 | $5 | 2012 | Apr08 | Web | | NewEgg.com | 8388608 | 40.99 | 9-9-9-24 | 2x4 GB DIMM DDR3-1333 @ $40.99 + free shipping | Patriot | |
| 2012.33 | $5 | 2012 | May24 | Web | | NewEgg.com | 8388608 | 39.99 | 9-9-9-24 | 2x4 GB DIMM DDR3-1333 @ $39.99 + free shipping | Team Elite | |
| 2012.58 | $5 | 2012 | Aug20 | Web | | NewEgg.com | 1.70E+07 | 77.99 | cas 9 | 2x8 GB DIMM DDR3-1333 @ $77.99 + free shipping | Kingston | |
| 2012.67 | $4 | 2012 | Sep21 | Web | | NewEgg.com | 1.70E+07 | 64.79 | 10-10-10-28 | 2x8 GB DIMM DDR3-1600 @ $64.79 + free shipping | GEIL EVO Corsa | |
| 2012.83 | $4 | 2012 | Nov17 | Web | | NewEgg.com | 8388608 | 29.99 | cas 9 | 2x4 GB DIMM DDR3-1333 @ $29.99 + free shipping | Crucial Ballistix | |
| 2013 | $4 | 2013 | Jan17 | Web | | NewEgg.com | 8388608 | 34.99 | 9-9-9-24 | 2x4 GB DIMM DDR3-1600 @ $34.99 + free shipping | Wintec One | |
| 2013.08 | $5 | 2013 | Feb23 | Web | | NewEgg.com | 1.70E+07 | 88.99 | 9-9-9-28 | 2x8 GB DIMM DDR3-1600 @ $88.99 + free shipping | GEIL EVO Corsa | |
| 2013.33 | $7 | 2013 | May18 | Web | | NewEgg.com | 1.70E+07 | 109.99 | 9-10-9-28 | 2x8 GB DIMM DDR3-1866 @ $109.99 + free shipping | GEIL | |
| 2013.42 | $6 | 2013 | Jun30 | Web | | NewEgg.com | 8388608 | 49.99 | cas 9 | 2x4 GB DIMM DDR3-1333 @ $49.99 + free shipping | | |
| 2013.58 | $7 | 2013 | Aug09 | Web | | NewEgg.com | 8388608 | 59.99 | 9-9-2009 | 2x4 GB DIMM DDR3-1333 @ $59.99 + free shipping | WINTEC value | |
| 2013.67 | $6 | 2013 | Sep07 | Web | | NewEgg.com | 8388608 | 52.99 | cas 10 | 1x8 GB DIMM DDR3-1600 @ $52.99 + free shipping | Crucial | |
| 2013.75 | $8 | 2013 | Oct16 | Web | | NewEgg.com | 1.70E+07 | 133.99 | 10-11-10-30 | 2x8 GB DIMM DDR3-1866 @ $133.99 + free shipping | Patriot Intel ExMLE | |
| 2013.83 | $9 | 2013 | Nov10 | Web | | NewEgg.com | 8388608 | 69.99 | cas 9 | 1x8 GB DIMM DDR3-1600 @ $69.99 + free shipping | Crucial Ballistix Sport | |
| 2013.92 | $8 | 2013 | Dec23 | Web | | NewEgg.com | 1.70E+07 | 129.99 | cas 11 | 2x8 GB DIMM DDR3-1600 @ $129.99 + free shipping | Silicon Power | |
| 2014.08 | $10 | 2014 | Feb08 | Web | | NewEgg.com | 4194304 | 38.99 | cas 10 | 1x4 GB DIMM DDR3-1600 @ $38.99 + free shipping | Patriot Viper 3 | |
| 2014.17 | $8 | 2014 | Mar13 | Web | | NewEgg.com | 8388608 | 64.99 | cas 9 | 2x4 GB DIMM DDR3-1600 @ $64.99 + free shipping | HyperX XMP | |
| 2014.25 | $7 | 2014 | Apr10 | Web | | NewEgg.com | 4194304 | 29.99 | 11-11-11-28 | 1x4 GB DIMM DDR3-1600 @ $29.99 + free shipping | Team | |
| 2014.42 | $8 | 2014 | Jun19 | Web | | NewEgg.com | 8388608 | 64.99 | 9-9-9-24 | 2x4 GB DIMM DDR3-1333 @ $64.99 + free shipping | Team Elite | |
| 2014.58 | $9 | 2014 | Aug02 | Web | | NewEgg.com | 8388608 | 69.99 | 9-9-9-24 | 2x4 GB DIMM DDR3-1600 @ $69.99 + free shipping | Team Vulcan | |
| 2014.67 | $9 | 2014 | Sep13 | Web | | NewEgg.com | 8388608 | 69.99 | 9-9-9-24 | 2x4 GB DIMM DDR3-1600 @ $69.99 + free shipping | Team Vulcan | |
| 2014.83 | $9 | 2014 | Nov11 | Web | | NewEgg.com | 4194304 | 34.99 | 11-11-11-30 | 1x4 GB DIMM DDR3-1600 @ $34.99 + free shipping | Pareema | |
| 2015 | $8 | 2015 | Jan22 | Web | | NewEgg.com | 8388608 | 63.99 | cas 10 | 1x8 GB DIMM DDR3-1866 @ $63.99 + free shipping | Hypexx Fury | |
| 2015.08 | $7 | 2015 | Feb15 | Web | | NewEgg.com | 8388608 | 59.99 | cas 11 | 2x4 GB DIMM DDR3-1600 @ $59.99 + free shipping | Crucial | |
| 2015.25 | $6 | 2015 | Apr11 | Web | | NewEgg.com | 8388608 | 49.99 | 11-11-11-28 | 2x4 GB DIMM DDR3-1600 @ $49.99 + free shipping | Avexir | |
| 2015.33 | $6 | 2015 | May15 | Web | | NewEgg.com | 1.70E+07 | 91.99 | 9-9-9-24 | 2x8 GB DIMM DDR3-1600 @ $91.99 + free shipping | Team Vulcan | |
| 2015.5 | $5 | 2015 | Jul10 | Web | | NewEgg.com | 8388608 | 39.99 | 11-11-11-28 | 1x8 GB DIMM DDR3-1600 @ $39.99 + free shipping | Team Elite | |
| 2015.58 | $5 | 2015 | Aug23 | Web | | NewEgg.com | 1.70E+07 | 73.99 | 9-9-9-24 | 2x8 GB DIMM DDR3L-1600 @ $73.99 + free shipping | Mushkin Enhanced ECO2 | |
| 2015.67 | $4 | 2015 | Sep22 | Web | | NewEgg.com | 8388608 | 34.99 | 9-9-9-24 | 2x4 GB DIMM DDR3-1333 @ $34.99 + free shipping | Team Elite | |
| 2015.75 | $4 | 2015 | Oct24 | Web | | NewEgg.com | 1.70E+07 | 68.99 | 9-9-9-24 | 2x8 GB DIMM DDR3-1600 @ $68.99 + free shipping | Team Dark | |
| 2015.83 | $4 | 2015 | Nov21 | Web | | NewEgg.com | 1.70E+07 | 62.99 | 9-9-9-24 | 2x8 GB DIMM DDR3-1600 @ $62.99 + free shipping | V-Color OC | |
| 2015.92 | $4 | 2015 | Dec18 | Web | | NewEgg.com | 1.70E+07 | 59.99 | 10-10-10-27 | 2x8 GB DIMM DDR3-1600 @ $59.99 + free shipping | Team Vulcan | |
| 2016.08 | $4 | 2016 | Feb14 | Web | | NewEgg.com | 1.70E+07 | 58.99 | 9-9-9-24 | 2x8 GB DIMM DDR3L-1600 @ $58.99 + free shipping | Mushkin Ehanced Essentials | |
| 2016.25 | $3 | 2016 | Apr11 | Web | | NewEgg.com | 1.70E+07 | 49.99 | 9-9-9-24 | 2x8 GB DIMM DDR3L-1600 @ $49.99 + free shipping | Mushkin Ehanced Essentials | |
| 2016.33 | $3 | 2016 | May18 | Web | | NewEgg.com | 1.70E+07 | 48.99 | 15-15-15-35 | 2x8 GB DIMM DDR4-2133 @ $48.99 + free shipping | Team Elite Plus | |
| 2016.42 | $3 | 2016 | Jun20 | Web | | NewEgg.com | 1.70E+07 | 44.99 | 10-11-10-30 | 2x8 GB DIMM DDR3-1866 @ $44.99 + free shipping | V-Color OC Series | |
| 2016.5 | $3 | 2016 | Jul22 | Web | | NewEgg.com | 1.70E+07 | 49.99 | 9-9-9-24 | 2x8 GB DIMM DDR3-1600 @ $49.99 + free shipping | Team Vulcan | |
| 2016.58 | $4 | 2016 | Aug22 | Web | | NewEgg.com | 1.70E+07 | 57.99 | 9-9-9-24 | 2x8 GB DIMM DDR3-1600 @ $57.99 + free shipping | Team Vulcan | |
| 2016.75 | $4 | 2016 | Oct15 | Web | | NewEgg.com | 3.40E+07 | 114.99 | 15-15-15-35 | 2x16 GB DIMM DDR4-2133 @ $114.99 + free shipping | G. Skill Aegis | |
| 2016.83 | $4 | 2016 | Nov15 | Web | | NewEgg.com | 1.70E+07 | 67.99 | 15-15-15-36 | 2x8 GB DIMM DDR4-2133 @ $67.99 + free shipping | GEIL EVO Potenza | |
| 2016.92 | $5 | 2016 | Dec21 | Web | | NewEgg.com | 1.70E+07 | 74.99 | 16-16-16-36 | 2x8 GB DIMM DDR4-3000 @ $74.99 + free shipping | Team Dark | |
| 2017 | $5 | 2017 | Jan15 | Web | | NewEgg.com | 1.70E+07 | 79.99 | 16-18-18-38 | 2x8 GB DIMM DDR4-3000 @ $79.99 + free shipping | Team Vulcan | |
| 2017.17 | $5 | 2017 | Mar12 | Web | | NewEgg.com | 1.70E+07 | 89.99 | 11-11-11-27 | 2x8 GB DIMM DDR3-2133 @ $89.99 + free shipping | Ballistic Elite | |
| 2017.33 | $6 | 2017 | May05 | Web | | NewEgg.com | 3.40E+07 | 185.99 | 15-15-15-36 | 2x16 GB DIMM DDR4-2133 @ $185.99 + free shipping | GEIL Forza | |
| 2017.42 | $5 | 2017 | Jun22 | Web | | NewEgg.com | 1.70E+07 | 89.99 | 9-9-9-24 | 2x8 GB DIMM DDR3-1600 @ $89.99 + free shipping | Team Dark | |
| 2017.5 | $6 | 2017 | Jul22 | Web | | NewEgg.com | 1.70E+07 | 95.99 | 9-9-9-24 | 2x8 GB DIMM DDR3-1600 @ $95.99 + free shipping | Adata XPG V1.0 | |
| 2017.67 | $6 | 2017 | Sep19 | Web | | NewEgg.com | 8388608 | 49.99 | 9-9-9-24 | 2x4 GB DIMM DDR3-1600 @ $49.99 + free shipping | Team Vulcan | |
| 2017.75 | $7 | 2017 | Oct27 | Web | | NewEgg.com | 1.70E+07 | 109.99 | | 2x8 GB DIMM DDR3-1600 @ $109.99 + free shipping | Team Elite Plus | |
| 2017.92 | $7 | 2017 | Dec12 | Web | | NewEgg.com | 1.70E+07 | 115.99 | | 2x8 GB DIMM DDR3-1600 @ $115.99 + free shipping | Silicon Power | |
| 2018.08 | $7 | 2018 | Feb14 | Web | | NewEgg.com | 1.70E+07 | 108.99 | | 2x8 GB DIMM DDR3-1600 @ $108.99 + free shipping | Silicon Power | |
| 2018.17 | $7 | 2018 | Mar23 | Web | | NewEgg.com | 1.70E+07 | 113.99 | | 2x8 GB DIMM DDR3-1600 @ $113.99 + free shipping | Team Dark | |
| 2018.33 | $7 | 2018 | May11 | Web | | NewEgg.com | 1.70E+07 | 111.98 | | 2x8 GB DIMM DDR3-1600 @ $110.99 + $0.99 shipping | Team Dark | |
| 2018.42 | $6 | 2018 | Jun19 | Web | | NewEgg.com | 1.70E+07 | 104.99 | 11-11-11-28 | 2x8 GB DIMM DDR3-1600 @ $104.99 + free shipping | Adata XPG V1.0 | |
| 2018.5 | $7 | 2018 | Jul08 | Web | | NewEgg.com | 1.70E+07 | 110.98 | 9-9-9-24 | 2x8 GB DIMM DDR3-1600 @ $109.99 + $0.99 shipping | Team Vulcan | |
| 2018.58 | $6 | 2018 | Aug24 | Web | | NewEgg.com | 1.70E+07 | 100.98 | | 2x8 GB DIMM DDR3-1600 @ $99.99 + $0.99 shipping | Team Dark | |
| 2018.67 | $5 | 2018 | Sep20 | Web | | NewEgg.com | 1.70E+07 | 89.99 | 11-11-11-28 | 2x8 GB DIMM DDR3-1600 @ $89.99 + free shipping | G.Skill | |
| 2018.83 | $5 | 2018 | Nov11 | Web | | NewEgg.com | 1.70E+07 | 82.99 | 11-11-11-28 | 2x8 GB DIMM DDR3L-1600 @ $82.99 + free shipping | Mushkin Enhanced Essentials | |
| 2018.92 | $5 | 2018 | Dec16 | Web | | NewEgg.com | 8388608 | 37.99 | | 1x8 GB DIMM DDR3-1333 @ $37.99 + free shipping | G.Skill Value Series | |
| 2019 | $4 | 2019 | Jan17 | Web | | NewEgg.com | 1.70E+07 | 69.99 | 11-11-11-28 | 2x8 GB DIMM DDR3L-1600 @ $69.99 + free shipping | Mushkin Enhanced Essentials | |
| 2019.08 | $4 | 2019 | Feb11 | Web | | NewEgg.com | 1.70E+07 | 69.99 | 11-11-11-28 | 2x8 GB DIMM DDR3L-1600 @ $69.99 + free shipping | Mushkin Enhanced Essentials | |
| 2019.17 | $4 | 2019 | Mar17 | Web | | NewEgg.com | 1.70E+07 | 68.99 | 11-11-11-28 | 2x8 GB DIMM DDR3L-1600 @ $68.99 + free shipping | Mushkin Enhanced Essentials | |
| 2019.25 | $4 | 2019 | Apr21 | Web | | NewEgg.com | 1.70E+07 | 60.98 | 11-11-11-28 | 2x8 GB DIMM DDR3-1600 @ $59.99 + $0.99 shipping | G.Skill Value | |
| 2019.33 | $3 | 2019 | 20-May | Web | | NewEgg.com | 8388608 | 27.99 | CL 11 | 1x 8GB DIMM DDR3-1600 @ $27.99 + free shipping | Patriot Signature Line | |
| 2019.42 | $3 | 2019 | 19-Jun | Web | | NewEgg.com | 8388608 | 25.99 | 17-17-17-39 | 1x 8GB DIMM DDR4-2400 @ $25.99 + free shipping | OLOy | |
| 2019.5 | $3 | 2019 | 18-Jul | Web | | NewEgg.com | 16777216 | 44.99 | 17-17-17-39 | 2x 8GB DIMM DDR4-2400 @ $44.99 + free shipping | OLOy | |
| 2019.58 | $3 | 2019 | 17-Aug | Web | | NewEgg.com | 16777216 | 44.99 | 17-17-17-39 | 2x 8GB DIMM DDR4-2400 @ $44.99 + free shipping | OLOy | |
| 2019.67 | $3 | 2019 | 16-Sep | Web | | NewEgg.com | 16777216 | 53.99 | 16-18-18-35 | 2x 8GB DIMM DDR4-2666 @ $53.99 + free shipping | OLOy | |
| 2019.75 | $3 | 2019 | 18-Oct | Web | | NewEgg.com | 33554432 | 99.99 | 19-19-19-43 | 2x 16GB DIMM DDR4-2666 @ $99.99 + free shipping | G. Skill Aegis | |
| 2019.83 | $3 | 2019 | 14-Nov | Web | | NewEgg.com | 33554432 | 94.99 | 16-18-18-36 | 2x 16GB DIMM DDR4-3000 @ $94.99 + free shipping | Geil EVO Potenza | |
| 2019.92 | $3 | 2019 | 17-Dec | Web | | NewEgg.com | 33554432 | 92.99 | 16-18-18-38 | 2x 16GB DIMM DDR4-3000 @ $92.99 + free shipping | OLOy | |
| 2020 | $3 | 2020 | 20-Jan | Web | | NewEgg.com | 16777216 | 47.99 | 17-17-17-39 | 2x 8GB DIMM DDR4-2400 @ $47.99 + free shipping | OLOy | |
| 2020.08 | $3.17 | 2020 | 17-Feb | Web | | NewEgg.com | 33554432 | 99.99 | 17-17-17-39 | 2x 16GB DIMM DDR4-2400 @ $99.99 + free shipping | OLOy | |
| 2020.17 | $3.58 | 2020 | 23-Mar | Web | | NewEgg.com | 33554432 | 114.99 | 17-17-17-39 | 2x 16GB DIMM DDR4-2400 @ $114.99 + free shipping | OLOy | |
| 2020.25 | $3.58 | 2020 | 23-Apr | Web | | NewEgg.com | 16777216 | 57.99 | 16-18-18-36 | 1x 16GB DIMM DDR4-3000 @ $57.99 + free shipping | OLOy | |
| 2020.33 | $3.58 | 2020 | 23-May | Web | | NewEgg.com | 33554432 | 114.99 | 19-19-19-43 | 2x 16GB DIMM DDR4-2666 @ $114.99 + free shipping | G. Skill | |
| 2020.42 | $3.28 | 2020 | 21-Jun | Web | | NewEgg.com | 33554432 | 104.99 | 15-15-15-39 | 2x 16GB DIMM DDR4-2400 @ $104.99 + free shipping | G. Skill X series | |
| 2020.5 | $2.97 | 2020 | 18-Jul | Web | | NewEgg.com | 33554432 | 94.99 | 19-19-19-43 | 2x 16GB DIMM DDR4-2666 @ $94.99 + free shipping | G. Skill Aegis | |
| 2020.58 | $2.76 | 2020 | 19-Aug | Web | | NewEgg.com | 33554432 | 89.99 | 22-22-22-52 | 1x 32GB DIMM DDR4-3200 @ $89.99 + free shipping | Team Elite | |
| 2020.67 | $2.87 | 2020 | 20-Sep | Web | | NewEgg.com | 33554432 | 91.99 | 16-18-18-36 | 2x 16GB DIMM DDR4-2666 @ $91.99 + free shipping | OLOy | |
| 2020.75 | $2.76 | 2020 | 17-Oct | Web | | NewEgg.com | 33554432 | 89.99 | 16-18-18-36 | 2x 16GB DIMM DDR4-3000 @ $89.99 + free shipping | GEIL Evo Potenza | |
| 2020.83 | $2.97 | 2020 | 19-Nov | Web | | NewEgg.com | 33554432 | 94.99 | 20-22-22-46 | 2x 16GB DIMM DDR4-3200 @ $94.99 + free shipping | Team T-Force | |
| 2020.92 | $3.07 | 2020 | 21-Dec | Web | | NewEgg.com | 33554432 | 97.99 | 16-18-18-38 | 2x 16GB DIMM DDR4-3000 @ $97.99 + free shipping | OLOy | |
| 2021.08 | $3.48 | 2021 | 14-Feb | Web | | NewEgg.com | 33554432 | 109.99 | 16-18-18-38 | 2x 16GB DIMM DDR4-3000 @ $109.99 + free shipping | G.Skill Aegis | |
| 2021.25 | $4.10 | 2021 | 17-Apr | Web | | NewEgg.com | 67108864 | 259.99 | 19-19-19-43 | 2x 32GB DIMM DDR4-2666 @ $259.99 + free shipping | G.Skill Value Series | |
| 2021.33 | $3.58 | 2021 | 17-May | Web | | NewEgg.com | 67108864 | 229.99 | 19-19-19-43 | 2x 32GB DIMM DDR4-2666 @ $229.99 + free shipping | G.Skill Value Series | |
| 2021.5 | $3.99 | 2021 | 11-Jul | Web | | NewEgg.com | 8388608 | 31.99 | cas 11 | 1x 8GB DIMM DDR3-1600 @ $31.99 + free shipping | Patriot Signature Line | |
| 2021.58 | $3.79 | 2021 | 16-Aug | Web | | NewEgg.com | 33554432 | 119.99 | 15-15-15-36 | 2x 16GB DIMM DDR4-2133 @ $119.99 + free shipping | G.Skill Ripjaws V | |
| 2021.67 | $3.28 | 2021 | 18-Sep | Web | | NewEgg.com | 16777216 | 51.99 | 19-19-19-43 | 1x 16GB DIMM DDR4-2666 @ $51.99 + free shipping | G.Skill Aegis | |
| 2021.75 | $3.07 | 2021 | 20-Oct | Web | | NewEgg.com | 16777216 | 48.99 | 17-17-17-39 | 1x 16GB DIMM DDR4-2400 @ $48.99 + free shipping | G.Skill Aegis | |
| 2021.92 | $2.66 | 2021 | 15-Dec | Web | | NewEgg.com | 33554432 | 84.99 | 19-19-19-43 | 2x 16GB DIMM DDR4-2666 @ $84.99 + free shipping | G.Skill Aegis | |
| 2022 | $2.87 | 2022 | 19-Jan | Web | | NewEgg.com | 33554432 | 92.99 | 19-19-19-43 | 2x 16GB DIMM DDR4-2666 @ $92.99 + free shipping | G.Skill Aegis | |
| 2022.08 | $2.97 | 2022 | 17-Feb | Web | | NewEgg.com | 33554432 | 93.99 | 19-19-19-43 | 2x 16GB DIMM DDR4-2666 @ $93.99 + free shipping | G.Skill Aegis | |
| 2022.17 | $3.17 | 2022 | 18-Mar | Web | | NewEgg.com | 33554432 | 99.99 | 16-20-20-40 | 2x 16GB DIMM DDR4-3200 @ $99.99 + free shipping | Team T-Force Vulcan TUF G.A. | |
| 2022.25 | $3.17 | 2022 | 23-Apr | Web | | NewEgg.com | 16777216 | 50.98 | 17-17-17-39 | 1x 16GB DIMM DDR4-2400 @ $49.99 + $0.99 shipping | G.Skill Aegis | |
| 2022.33 | $2.76 | 2022 | 15-May | Web | | NewEgg.com | 33554432 | 87.99 | cas 22 | 1x 32GB SO-DIMM DDR4-3200 @ $87.99 + free shipping | G.Skill Ripjaws | |
| 2022.42 | $2.66 | 2022 | 18-Jun | Web | | NewEgg.com | 33554432 | 84.97 | cas 22 | 2x 16GB SO-DIMM DDR4-3200 @ $84.97 + free shipping | Silicon Power | |
| 2022.58 | $2.36 | 2022 | 20-Aug | Web | | NewEgg.com | 33554432 | 74.99 | cas 22 | 1x 32GB SO-DIMM DDR4-3200 @ $74.99 + free shipping | OLOy | |
| 2022.67 | $2.25 | 2022 | 16-Sep | Web | | NewEgg.com | 33554432 | 72.99 | cas 19 | 1x 32GB SO-DIMM DDR4-2666 @ $72.99 + free shipping | Silicon Power | |
| 2022.75 | $2.25 | 2022 | 18-Oct | Web | | NewEgg.com | 33554432 | 72.99 | cas 19 | 2x 16GB SO-DIMM DDR4-2666 @ $72.99 + free shipping | Silicon Power | |
| 2022.83 | $2.15 | 2022 | 5-Nov | Web | | NewEgg.com | 33554432 | 69.99 | cas 19 | 1x 32GB SO-DIMM DDR4-2666 @ $69.99 + free shipping | Silicon Power | |
| 2022.92 | $2.15 | 2022 | 18-Dec | Web | | NewEgg.com | 33554432 | 68.99 | 17-17-17-39 | 2x 16GB DIMM DDR4-2400 @ $68.99 + free shipping | G.Skill Aegis | |
| 2023 | $2.05 | 2023 | 18-Jan | Web | | NewEgg.com | 33554432 | 63.98 | cas 19 | 1x 32GB SO-DIMM DDR4-2666 @ $61.99 + $1.99 shipping | G.Skill Ripjaws | |
| 2023.08 | $1.84 | 2023 | 18-Feb | Web | | NewEgg.com | 67108864 | 114.99 | cas 22 | 2x 32GB SO-DIMM DDR4-3200 @ $114.99 + free shipping | Mushkin Enhanced Essentials | |
| 2023.17 | $1.84 | 2023 | 18-Mar | Web | | NewEgg.com | 67108864 | 114.99 | 22-22-22-52 | 2x 32GB SO-DIMM DDR4-3200 @ $114.99 + free shipping | Mushkin Enhanced Essentials | |
| 2023.33 | $1.54 | 2023 | 6-May | Web | | NewEgg.com | 67108864 | 97.99 | cas 22 | 2x 32GB SO-DIMM DDR4-3200 @ $97.99 + free shipping | Team Elite | |
| 2023.42 | $1.46 | 2023 | 17-Jun | Web | | NewEgg.com | 67108864 | 93.99 | 19-19-19-43 | 2x 32GB DIMM DDR4-2666 @ $93.99 + free shipping | Mushkin Enhanced Essentials | |
| 2023.5 | $1.43 | 2023 | 18-Jul | Web | | NewEgg.com | 33554432 | 45.99 | | 2x 16GB SO-DIMM DDR4-3200 @ $45.99 + free shipping | Team T-Create Classic | |
| 2023.67 | $1.45 | 2023 | 19-Sep | Web | | NewEgg.com | 67108864 | 92.99 | 16-19-19-38 | 2x 32GB DIMM DDR4-3200 @ $92.99 + free shipping | Mushkin Enhanced Redline Stiletto | |
| 2023.75 | $1.40 | 2023 | 20-Oct | Web | | NewEgg.com | 67108864 | 89.99 | 22-22-22-52 | 2x 32GB SO-DIMM DDR4-3200 @ $89.99 + free shipping | Team T-Create Classic | |
| 2023.83 | $1.31 | 2023 | 16-Nov | Web | | NewEgg.com | 33554432 | 41.99 | 16-17-17-36 | 2x 16GB SO-DIMM DDR4-2666 @ $41.99 + free shipping | Mushkin Enhanced Redline | |
| 2024.00 | $1.61 | 2024 | 14-Jan | Web | | NewEgg.com | 67108864 | 102.99 | 22-22-22-52 | 2x32 GB SO-DIMM DDR4-3200 @102.99 + free shipping | Team Elite | |
| 2024.08 | $1.72 | 2024 | 21-Feb | Web | | NewEgg.com | 67108864 | 109.99 | 22-22-22-52 | 2x32 GB SO-DIMM DDR4-3200 @109.99 + free shipping | Mushkin Enhanced Essentials | |
| 2024.17 | $1.57 | 2024 | 21-Mar | Web | | NewEgg.com | 67108864 | 99.99 | 16-18-18-35 | 2x32 GB DIMM DDR4-2666 @99.99 + free shipping | Corsair Vengence LPX | |
| 2024.33 | $1.66 | 2024 | 16-May | Web | | NewEgg.com | 33554432 | 52.99 | 22-22-22-52 | 2x16 GB DIMM DDR4-3200 @ $52.99 + free shipping | Team Elite | |
| 2024.33 | $3.12 | 2024 | 28-Jul | Web | | NewEgg.com | 33554432 | 99.99 | 20-26-26-46 | 2x32 GB DIMM DDR4-3600 @99.99 + free shipping | Patriot Viper Elite | |
| 2024.33 | $1.57 | 2025 | 15-Apr | Web | | NewEgg.com | 33554432 | 49.99 | 15-20-20 | Kingston FURY Beast 32GB DDR4 3200MHz DIMM Memory Moc | kingston furry | |